\documentclass[acmsmall]{acmart}

\usepackage[utf8]{inputenc}
\usepackage[T1]{fontenc}
\usepackage{hyphenat}
\usepackage{xspace}
\usepackage{amsmath}
\usepackage{amsfonts}
\usepackage{hyperref}
\usepackage{url}
\usepackage{multirow}
\usepackage{makecell}
\usepackage{caption}
\usepackage{minibox}
\usepackage{bbm}
\usepackage{graphicx}
\usepackage{balance}
\usepackage{mathtools}
\usepackage{color}
\usepackage{marvosym}
\usepackage{ifthen}
\usepackage{textcomp}
\usepackage{enumitem}
\usepackage{verbatim}
\usepackage{algorithm}
\usepackage{algorithmic}
\usepackage{numprint}
\usepackage{balance}

\usepackage{subcaption}

\usepackage{amsthm}
\theoremstyle{plain}

\newcommand{\chatoDisplayMode}[1]{#1}

\definecolor{MyRed}{rgb}{0.6,0.0,0.0} 
\definecolor{MyBlack}{rgb}{0.1,0.1,0.1} 
\newcommand{\inred}[1]{{\color{MyRed}\sf\textbf{\textsc{#1}}}}
\newcommand{\frameit}[2]{
  \begin{center}
  {\color{MyRed}
  \framebox[.9\columnwidth][l]{
    \begin{minipage}{.85\columnwidth}
    \inred{#1}: {\sf\color{MyBlack}#2}
    \end{minipage}
  }\\
  }
  \end{center}
}

\newcommand{\note}[2][]{\chatoDisplayMode{\def\@tmpsig{#1}\frameit{{\Pointinghand} Note}{#2\ifx \@tmpsig \@empty \else \mbox{ --\em #1}\fi}}}
\newcommand{\todo}[2][]{\chatoDisplayMode{\def\@tmpsig{#1}\frameit{{\Writinghand} To-do}{#2\ifx \@tmpsig \@empty \else \mbox{ --\em #1}\fi}}}

\newcommand{\abbrevStyle}[1]{#1}

\newcommand{\ie}{\abbrevStyle{i.e.}\xspace}
\newcommand{\eg}{\abbrevStyle{e.g.}\xspace}
\newcommand{\cf}{\abbrevStyle{cf.}\xspace}

\newcommand{\vs}{\abbrevStyle{vs.}\xspace}

\newcommand{\Eqnref}[1]{Eq.~\ref{#1}}

\newcommand{\xhdr}[1]{\vspace{1.7mm}\noindent{{\bf #1.}}}

\newcommand{\textcite}[1]{\citeauthor{#1} \shortcite{#1}}

\newcommand{\hide}[1]{}

\makeatletter
\newcommand{\iffont}[2]{\ifthenelse{\equal{\f@family}{#1}}{#2}{}}
\makeatother

\iffont{ptm}{
  \usepackage{mathptmx}

  \DeclareSymbolFont{greek}{OML}{cmm}{m}{n}
  \DeclareMathSymbol{\alpha}{\mathalpha}{greek}{"0B}
  \DeclareMathSymbol{\beta}{\mathalpha}{greek}{"0C}
  \DeclareMathSymbol{\gamma}{\mathalpha}{greek}{"0D}
  \DeclareMathSymbol{\delta}{\mathalpha}{greek}{"0E}
  \DeclareMathSymbol{\epsilon}{\mathalpha}{greek}{"0F}
  \DeclareMathSymbol{\zeta}{\mathalpha}{greek}{"10}
  \DeclareMathSymbol{\eta}{\mathalpha}{greek}{"11}
  \DeclareMathSymbol{\theta}{\mathalpha}{greek}{"12}
  \DeclareMathSymbol{\iota}{\mathalpha}{greek}{"13}
  \DeclareMathSymbol{\kappa}{\mathalpha}{greek}{"14}
  \DeclareMathSymbol{\lambda}{\mathalpha}{greek}{"15}
  \DeclareMathSymbol{\mu}{\mathalpha}{greek}{"16}
  \DeclareMathSymbol{\nu}{\mathalpha}{greek}{"17}
  \DeclareMathSymbol{\xi}{\mathalpha}{greek}{"18}
  \DeclareMathSymbol{\pi}{\mathalpha}{greek}{"19}
  \DeclareMathSymbol{\rho}{\mathalpha}{greek}{"1A}
  \DeclareMathSymbol{\sigma}{\mathalpha}{greek}{"1B}
  \DeclareMathSymbol{\tau}{\mathalpha}{greek}{"1C}
  \DeclareMathSymbol{\upsilon}{\mathalpha}{greek}{"1D}
  \DeclareMathSymbol{\phi}{\mathalpha}{greek}{"1E}
  \DeclareMathSymbol{\chi}{\mathalpha}{greek}{"1F}
  \DeclareMathSymbol{\psi}{\mathalpha}{greek}{"20}
  \DeclareMathSymbol{\omega}{\mathalpha}{greek}{"21}
  \DeclareMathSymbol{\varepsilon}{\mathalpha}{greek}{"22}
  \DeclareMathSymbol{\vartheta}{\mathalpha}{greek}{"23}
  \DeclareMathSymbol{\varpi}{\mathalpha}{greek}{"24}
  \DeclareMathSymbol{\varrho}{\mathalpha}{greek}{"25}
  \DeclareMathSymbol{\varsigma}{\mathalpha}{greek}{"26}
  \DeclareMathSymbol{\varphi}{\mathalpha}{greek}{"27}
  \DeclareSymbolFont{otone}{OT1}{cmr}{m}{n}
  \DeclareMathSymbol{\Gamma}{\mathalpha}{otone}{0}
  \DeclareMathSymbol{\Delta}{\mathalpha}{otone}{1}
  \DeclareMathSymbol{\Theta}{\mathalpha}{otone}{2}
  \DeclareMathSymbol{\Lambda}{\mathalpha}{otone}{3}
  \DeclareMathSymbol{\Xi}{\mathalpha}{otone}{4}
  \DeclareMathSymbol{\Pi}{\mathalpha}{otone}{5}
  \DeclareMathSymbol{\Sigma}{\mathalpha}{otone}{6}
  \DeclareMathSymbol{\Upsilon}{\mathalpha}{otone}{7}
  \DeclareMathSymbol{\Phi}{\mathalpha}{otone}{8}
  \DeclareMathSymbol{\Psi}{\mathalpha}{otone}{9}
  \DeclareMathSymbol{\Omega}{\mathalpha}{otone}{10}
  \DeclareSymbolFont{syms}{OML}{cmm}{m}{it}
  \DeclareMathSymbol{\partial}{\mathord}{syms}{"40}
  \DeclareMathAlphabet{\mathbold}{OML}{cmm}{b}{it}
  \DeclareSymbolFont{largesymbols}{OMX}{cmex}{m}{n}

}

\newcommand{\new}[1]{\textcolor{black}{#1}} 

\usepackage{wrapfig}
\begin{document}

\setcopyright{acmlicensed}
\acmJournal{PACMHCI}
\acmYear{2021}\acmVolume{5} \acmNumber{CSCW1} \acmArticle{184} \acmMonth{4} \acmPrice{15.00}\acmDOI{10.1145/3449297}

\copyrightyear{2021}

\received{October 2020} 
\received[revised]{January 2021}
\received[accepted]{January 2021}

\title[Formation of Social Ties Influences Food Choice: A Campus-Wide Longitudinal Study]{%
Formation of Social Ties Influences Food Choice:\\
A Campus-Wide Longitudinal Study}

\author{Kristina Gligori\'c}
\affiliation{%
  \institution{EPFL}
  \country{Lausanne, Switzerland}
}
\email{kristina.gligoric@epfl.ch}

\author{Ryen W. White}
\affiliation{%
  \institution{Microsoft Research}
  \country{Redmond, Washington, United States}
}
\email{ryenw@microsoft.com}

\author{Emre K{\i}c{\i}man}
\affiliation{%
  \institution{Microsoft Research}
  \country{Redmond, Washington, United States}
}
\email{emrek@microsoft.com}

\author{Eric Horvitz}
\affiliation{%
  \institution{Microsoft Research}
  \country{Redmond, Washington, United States}
}
\email{horvitz@microsoft.com}

\author{Arnaud Chiolero}
\authornote{Also affiliated with Institute of Primary Health Care (BIHAM), University of Bern, Bern, Switzerland, and School of Population and Global Health, McGill University, Montreal, Canada.}
\affiliation{%
  \institution{%
  Population Health Laboratory,
  University of Fribourg}
  \country{Fribourg, Switzerland}
}
\email{achiolero@gmail.com}

\author{Robert West}
\affiliation{%
  \institution{EPFL}
  \country{Lausanne, Switzerland}
}
\email{robert.west@epfl.ch}

\renewcommand{\shortauthors}{Gligori\'c et al.}

\begin{abstract}
Nutrition is a key determinant of long-term health, and social influence has long been theorized to be a key determinant of nutrition. It has been difficult to quantify the postulated role of social influence on nutrition using traditional methods such as surveys, due to the typically small scale and short duration of studies. To overcome these limitations, we leverage a novel source of data: logs of 38 million food purchases made over an 8-year period on the \'Ecole Polytechnique F\'ed\'erale de Lausanne (EPFL) university campus, linked to anonymized individuals via the smartcards used to make on-campus purchases. In a longitudinal observational study, we ask: How is a person's food choice affected by eating with someone else whose own food choice is healthy vs. unhealthy? To estimate causal effects from the passively observed log data, we control confounds in a matched quasi-experimental design: we identify focal users who at first do not have any regular eating partners but then start eating with a fixed partner regularly, and we match focal users into comparison pairs such that paired users are nearly identical with respect to covariates measured before acquiring the partner, where the two focal users' new eating partners diverge in the healthiness of their respective food choice. A difference-in-differences analysis of the paired data yields clear evidence of social influence: focal users acquiring a healthy-eating partner change their habits significantly more toward healthy foods than focal users acquiring an unhealthy-eating partner. We further identify foods whose purchase frequency is impacted significantly by the eating partner's healthiness of food choice. Beyond the main results, the work demonstrates the utility of passively sensed food purchase logs for deriving insights, with the potential of informing the design of public health interventions and food offerings, especially on university campuses.
\end{abstract}


\keywords{nutrition; behavioral analysis; causal analysis; university; public health}

\maketitle

\section{Introduction}
\label{sec:intro}
Nutrition plays a key role in people's daily lives and is a major determinant of long-term health~\cite{delaney2011food,gakidou2017global}.
Interventions and policies that promote healthier diets are therefore important public health priorities \cite{willett2019food}.
In designing such interventions, situational food norms, including social influences, play a prominent role, as they are theorized to have a powerful effect on food intake~\cite{hetherington2006situational,higgs2016social,shepherd_1999,collins2019two,mollen2013healthy}.

Despite the postulated importance of social factors, measuring how dietary behaviors are affected by the behaviors of others remains challenging. 
On the one hand, experimental studies to date have been limited to observing people in small-scale scenarios with a short duration \cite{ROBINSON2014414,robinson2013food}.
On the other hand, observational studies have relied on survey-based methods \cite{christakis2007spread}, employing questionnaires \cite{wouters2010peer} and personal food journals \cite{barriers_negative2015,food_journal2015}, which are costly to organize and prone to biases \cite{bowling2005mode}.

Furthermore, making causal inferences regarding the influence of social ties on food intake faces the challenge of numerous confounding factors. Although similarities in diet and eating behaviors among persons connected via social links (\eg, friends, family and peers) have been observed in a number of experimental and survey‐based studies \cite{madan2010social,finnerty2010effects,patrick2005review,salvy2008effects,stevenson2007adolescents,harmon2016perceived}, it is not clear whether the similarity in the consumption patterns arises from social influence, or if confounding factors, such as self\hyp selection in tie formation (homophily) and environmental influences, can explain the similarity \cite{aral2009distinguishing,shalizi2011homophily,shalizi2016estimating,kossinets2006empirical}. In real-world settings, it remains challenging to measure and disentangle properties that are relevant in the context of food consumption, such as attributes of the individuals and of the environment (\eg, food options available in different locations and settings). Researchers have only recently been addressing this gap by studying social media and other digital traces of human behavior in the context of food consumption \cite{abbar2015you,hunger2016,pro_eating2016}.

In order to shed new light on the influence of social factors on food choice, we harness a novel data source:
logs of 38 million food purchases made over an eight-year period on a major university campus, linked to anonymized individuals (students and staff) via the smartcards they use to make on-campus purchases.
The large scale and long duration of the data enables studies with greater statistical power, compared to prior setups, and allows for reducing the influence of confounding factors, and thus for identifying the causal effect of social influence, by carefully selecting a suitable subset of individuals whose food choice behaviors are monitored throughout the individuals' long-term transaction histories.

Based on this dataset,
we design a longitudinal observational study to address the question of how a person's food choice is affected by eating with someone else whose own food choice is healthy vs. unhealthy.
To estimate causal effects from the passively observed log data, we control confounds in a matched quasi-experimental design.
where we identify focal users who at first do not have any regular eating partners but then start eating with a fixed partner regularly, and we match focal users into comparison pairs such that paired focal users are nearly identical with respect to covariates measured before acquiring the eating partner, but the two focal users' new eating partners diverge in the healthiness of their respective food choice.

\xhdr{Research questions}
Specifically, we seek to answer the following questions:

\begin{enumerate}
    \item How is the \textit{overall healthiness} of a focal person's food choice affected by the healthiness of an eating partner's food choice? Does the focal person's food choice change, and if so, in what direction?
    \item How is a focal person's choice of \textit{specific food categories} affected by the healthiness of an eating partner's food choice? Does the distribution over food categories change, and if so, what items are purchased more, and what items less?
\end{enumerate}

\xhdr{Summary of main findings}
Regarding question 1, we observe that, when a focal person acquires a new eating partner, the healthiness of the focal user's food choice shifts significantly in the direction of their new eating partner's dietary patterns.
In a difference-in-differences analysis of 415 comparison pairs of focal persons (identified among a total of around 39,000 persons in eight years' worth of log data), which carefully controls for a number of confounding covariates, we find clear evidence of social influence: focal persons acquiring a healthy-eating partner change their habits significantly more in the direction of healthy foods than focal persons acquiring an unhealthy-eating partner.
We quantify the robustness of this finding in a sensitivity analysis, and we provide further evidence by observing a dose--response relationship between the difference in exposures and the difference in effects.

Regarding question 2, we observe that focal persons who start eating with healthy-eating partners show an increase in the purchase of coffee and lunch meals, items generally purchased in large numbers, with the strongest effect. On the other hand, items purchased at higher rates by the matched counterparts, who start eating with unhealthy-eating partners, loosely form a cluster of potentially unhealthy items that should not be eaten in large quantities (soft drinks, drinks from vending machines, condiments, pizza, kebabs, and cr\^epes).

\xhdr{Implications}
Students and staff consume large amounts of food on campuses, daily and globally. The present work shows the value of employing novel methods relying on population-scale digital traces to measure social influence on food choice behaviors in this context. The derived insights have the potential to support interventions aimed at encouraging more healthy and sustainable dietary habits in university environments and beyond.

\section{Related work}
\label{sec:related}

\subsection{Social influence and diet}

Social influence on dietary habits is an active area of research \cite{higgs2016social,shepherd_1999}. Food consumption has been found to be influenced by eating with others \cite{hetherington2006situational}, and the food choices of others, including people one does not know, have been observed to influence food choices, even when not consciously recognized  \cite{christie2018vegetarian,robinson2013food}. Particular attention has been given to understanding the governing psychological mechanisms, including the seeking of dish uniformity driven by the goal of regret minimization, or the seeking of dish variety driven by self-presentation~\cite{ariely2000sequential,de2013adolescents,munt2017barriers}.

Although the underlying mechanisms are not fully understood, uniformity seeking is observed across a range of studies. For example, it is observed that the quantity dimension is used to communicate gender identity, and the food-type dimension to ingratiate the co-eater's preferences by matching the other's presumed choice, following gender-based stereotypes about food \cite{cavazza2017portion}. Such social norms, including the influence of peers, have tremendous potential for understanding dietary patterns and designing public health interventions \cite{collins2019two,mollen2013healthy,robinson_blissett_higgs_2013,ROBINSON2014414}.

In our work, we monitor behaviors outside of experimental setups. While previous efforts in this area have focused on specific behaviors (\eg, buying a dessert or not), having access to a multi-year history of all transactions made on a large campus allows us to observe behavioral changes for longer time periods and in a more fine-grained way, by measuring a wide set of purchasing behaviors that occur in the real world.

\subsection{The special case of children and adolescents}

A large fraction of the transactions recorded in our logs were made by students, \ie, adolescents and young adults.
Focusing on similar age groups, social influence in dietary habits has been examined in the context of school children \cite{finnerty2010effects,patrick2005review,salvy2008effects,birch1980effects} and adolescents \cite{stevenson2007adolescents,DELAHAYE2010161,DELAHAYE2011719}, who are theorized to be most susceptible to social pressures. In particular, effects of peer influence have been observed in children and adolescents' diets as well as activity patterns \cite{ball2010healthy,salvy2012influence}.

Systematic reviews of social network analyses of young people's eating behaviors and body weight reveal consistent evidence that school friends are significantly similar in terms of their body mass index. Friends with the highest body mass index appear to be most similar \cite{fletcher2011you}. Prior work further reveals that the family context is essential when implementing healthy eating interventions, as parents, not friends, are the most prominent influencers of adolescents' healthy eating \cite{eurpub,pedersen2015following}.

\subsection{Contagiousness of unhealthy behavior}

Previous work has particularly been focused on unhealthy behaviors and their contagious effects, observing that obesity \cite{christakis2007spread}, overeating \cite{doi:10.1086/644611}, fast food \cite{thornton2013barriers}, high-fat \cite{FEUNEKES1998645,hermans2009effects}, and alcohol and snack consumption \cite{pachucki2011social,wouters2010peer} are contagious. In fact, the strongest evidence of social influence in food choices has been found for unhealthy behaviors (\eg, snack foods) \cite{CRUWYS20153,blok2013unhealthy}. Beyond food consumption, peer influence and social norms have been observed to play a role in unhealthy weight-control behaviors among adolescent girls: self-induced vomiting, laxatives, diet pills, and fasting were all shown to be contagious among adolescent girls \cite{EISENBERG20051165}. A rich literature exists on tackling the problem of unhealthy behaviors through interventions with the goal of promoting healthy dietary habits and physical activity \cite{fjeldsoe2011systematic}, losing weight \cite{jeffery1993strengthening}, reducing the risk of chronic illnesses \cite{gittelsohn2012interventions}, and reducing food waste \cite{reynolds2019consumption}.

There is a heated debate about whether unhealthy behaviors are indeed contagious, or whether the observed similarities should instead be attributed to homophily, \ie, people's tendency to form ties with others who are similar to oneself to begin with.
Disentangling social influence from homophily poses a fundamental challenge. Without strong assumptions about the structure of ties or the ability to measure confounding factors, homophily and contagion are generically confounded (\ie, the effect of social influence cannot be identified) \cite{aral2009distinguishing,shalizi2011homophily,shalizi2016estimating}.

Our work attempts to minimize the effect of confounding variables in previously infeasible ways. Based on the rich transaction data, we measure a set of relevant confounding variables and carefully control for them in our quasi-experimental setup.

\subsection{Nutrition monitoring and modeling based on digital traces}

Social media has emerged as a promising source of data for studies on monitoring food consumption. For instance, it has been shown that Twitter has tremendous potential to provide insights into food choices at a population scale \cite{abbar2015you}. Researchers have also studied specific dietary issues and behaviors: reports of eating disorders \cite{hunger2016,pro_eating2016}, dietary choices, and nutritional challenges in food deserts, \ie, places with poor access to healthy and affordable food \cite{de2016characterizing}. Another active area of research has been focused on improving methods for monitoring food consumption, relying on mobile phones \cite{barriers_negative2015,food_journal2015} and wearable devices, to recognize when the eating activities take place~\cite{eating_moments}.

Recent related research has also demonstrated the value of monitoring and modeling of nutrition using other kinds of large-scale digital traces \cite{groseclose2017public}, such as grocery store purchase logs \cite{aiello2019large,buckeridge2014method}, online recipes \cite{rokicki2018impact}, logging-based smartphone applications and wearables \cite{achananuparp2016extracting,info:doi/10.2196/20625}, reviewing platforms~\cite{chorley2016pub}, search engine logs \cite{West:2013:CCI:2488388.2488510,vosen2011forecasting}, social media such as Twitter \cite{abbar2015you,mejova2016fetishizing,widener2014using} or Instagram~\cite{sharma2015measuring,ofli2017saki}, crowdsourcing platforms \cite{Howell:2016:ATP:2896338.2896358}, and geo-location signals \cite{sadilek2018machine}. 

Finally, large-scale passively sensed signals have been harnessed in university campus environments to measure factors of well-being outside of nutrition \cite{barclay2013peer,madan2010social,nook2015social,sefidgar2019passively,swain2020leveraging}. Recent preliminary insights point towards the feasibility and the potential of automatically inferring social interactions from behavioral traces for campus-centric applications \cite{swain2020leveraging}.

To summarize, while large-scale digital traces are promising for monitoring and modeling nutrition, little is known about how these passively sensed behavioral signals can be used for understanding the factors that govern food consumption in campus settings. Our longitudinal study aims to bridge this gap by analyzing large-scale, long-term purchase data.

\section{Materials and methods}
\label{sec:matmet}
\begin{figure}[t]
    \includegraphics[width=0.7\textwidth]{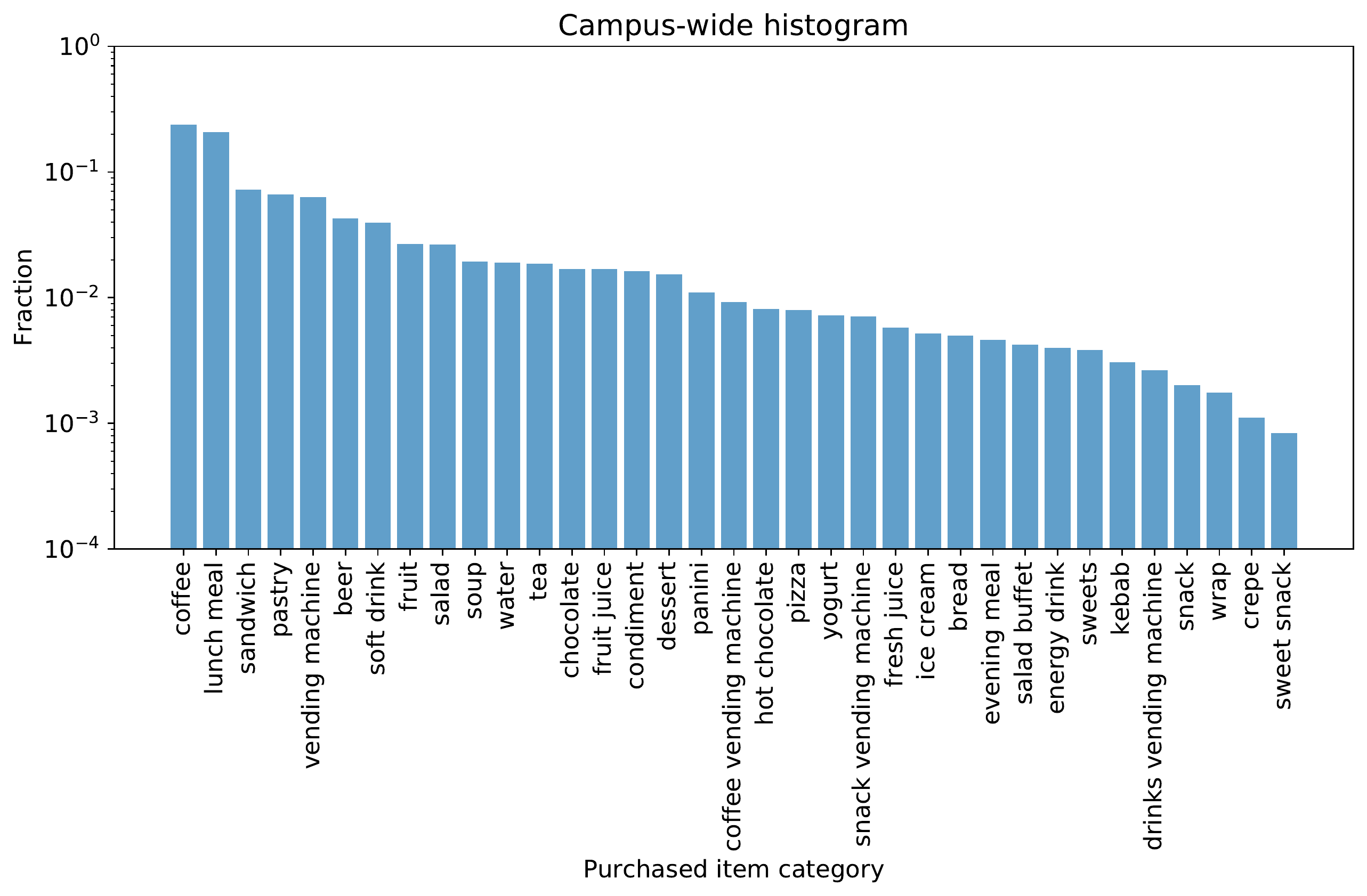}
    \caption{Distribution of purchases across item categories, across a campus. Most purchased items are coffee and lunch meals.}
    \label{fig:3}
\end{figure}

\subsection{Transaction log data}
\label{sec:Transaction log data}

This work leverages an anonymized dataset of food purchases made on the \'Ecole Polytechnique F\'ed\'erale de Lausanne (EPFL) university campus. The data spans 8 years, from 2010 to 2018, and contains about 38 million transactions, of which about 18 million were made with a badge that allows linking to an anonymized person's ID. The data includes 38.7k users, who, on median, are observed for a time period spanning 578 days and make 188 transactions. Each transaction is labeled with the time it took place, information about the sales location (shop, restaurant,  vending machine, or caf\'e), the cash register where the transaction took place, and the purchase items. Items are associated with unstructured textual descriptions (\eg, ``coffee'', ``croissant'', ``Coca-Cola can'').
The unstructured textual descriptions were additionally manually mapped to categorical labels (such as ``meal'', ``drink'', or ``dessert'') by a research assistant, who labeled the 500 most frequently purchased items, which account for 95.4\% of the total volume of item purchases observed in the dataset.
The distribution of purchases across categories is shown in Figure~\ref{fig:3}.

Purchases are not evenly spread over the course of the year, but, as expected, are higher during semesters, and lower during the breaks between semesters (Figure~\ref{fig:2}, left).

This work also leverages a smaller-size enriched transactional dataset gathered during a campus-wide sustainability challenge, for which 1,031 consenting participants formed 278 teams in order to compete in taking sustainable actions (\eg, taking the stairs instead of the elevator, or consuming a vegetarian meal). This data was not used for our analyses, but only for assessing the accuracy of our heuristic method for inferring frequent eating peers (described next).

\subsection{Inference of co-eating onset from proximity in transaction logs}
\label{sec:Inference of co-eating onset from proximity in transaction logs}

\begin{figure}[t]
        \includegraphics[width=\textwidth]{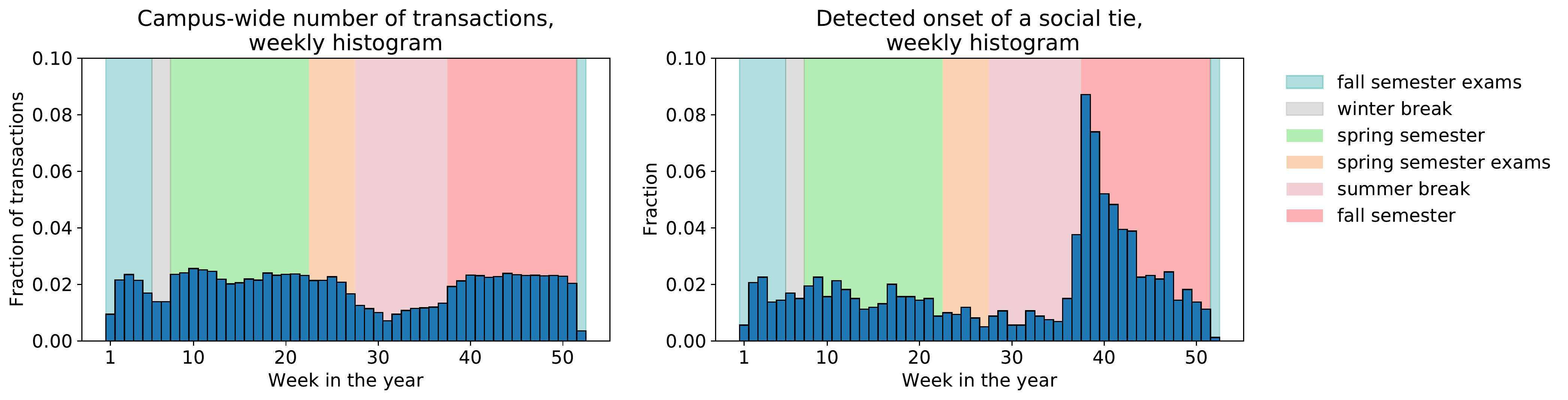}
    \caption{\textbf{Left:} Annual distribution of food purchases. The trends mirror the university schedule: the number of transactions drops at the end of the spring semester (around week 25), and increases again at the start of the fall semester (around week 40). A similar pattern is observed before the beginning of the spring semester (around week 10).
    \textbf{Right:} Annual distribution of detected onsets of social ties. The ties emerge disproportionally often when classes start at the beginning of the fall semester (by a factor of 3.5 times, compared to a baseline sampled at random from the distribution of purchases).}
    \label{fig:2}
\end{figure}

To measure the effect of the emergence of new social ties, we first infer frequently co-eating persons based on the proximity in the transaction logs. Frequently co-eating persons are likely to share a social tie, \ie, they are persons likely to be friends, colleagues, or classmates who often eat together. Previous work has shown that such social ties can be reliably inferred from geospatial proximity~\cite{crandall2010inferring}.
	
To infer frequent eating peers, we monitor a sequence of transactions made on the same day with the badge in the queue of a fixed cash registry, in a given shop. We identify situations when two individuals are adjacent in the queue and make a transaction within one minute between each other, with no one in between them. We use a lower threshold of 10 such high-confidence proximity indicators to infer a likely social tie. The first appearance of proximity in the logs is then considered to be the onset of co-eating. We observe a spike in tie formation coinciding with the start of classes in the fall (Figure~\ref{fig:2}, right).

Furthermore, we evaluate the precision of our heuristic by comparing the inferred co-eaters with ground-truth team membership information from the sustainability challenge. We observe that team membership in the sustainability challenge, a ground-truth indicator of a social tie, is correlated with sharing an inferred tie based on the transaction logs: out of all the pairs of individuals from the sub-population taking part in the sustainability challenge who are detected as frequent eating partners,
72\%
are also members of the same team.

\subsection{Inference of nutritional properties from raw transaction logs}

We infer a set of summary nutritional properties from raw transaction logs by relying on a set of pre-established criteria. We derive healthiness labels based on food-pyramid recommendations~\cite{walter2007food}. Products that should be consumed in the least amounts possible, \ie, items at the top of the Swiss food pyramid (with high amounts of saturated fats, salt, added sugars, refined grains, and highly processed foods) were considered as ``unhealthy'' (\eg, sodas, chips, candies, and chocolate bars). Other products that are not at the top of the Swiss food pyramid are considered to be ``healthy'' (including non-sweetened beverages, fruits, vegetables, whole grains, meat, fish, and nuts). When insufficient information was available from the name of the product, ``unclassifiable'' was selected.

Two professional epidemiologists specialized in nutrition independently assessed each food item and categorized them into healthy \vs\ unhealthy \vs\ unclassifiable. The reviewers had access to the unstructured textual description of the item (e.g., ``coffee'', ``croissant'', ``Coca-Cola can''). The reviewers did not have access to any other meta-information about the items. Disagreements were resolved by a third reviewer. Labels are used to create a healthiness score of a set of purchases by averaging individual product scores, coded numerically as $1$ for healthy (25\% of items), $-1$ for unhealthy (46\% of items), and $0$ for unclassifiable (29\% of items).

\subsection{Matched incident user design with active comparators}

\begin{figure}[t]
    \includegraphics[width = \textwidth]{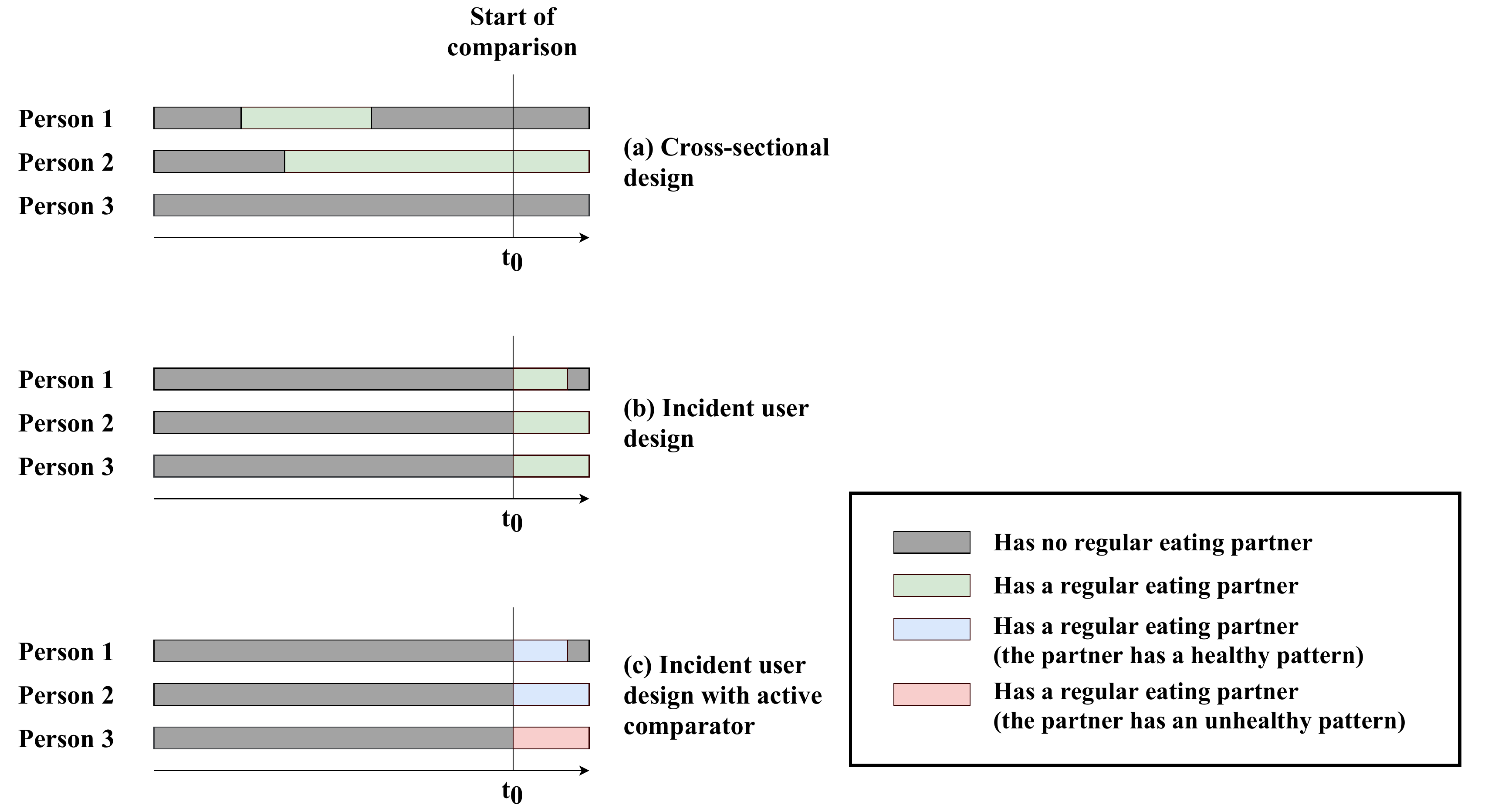}
    \caption{Study design diagrams. We illustrate three potential observational study designs to estimate the effect of eating with other persons on food choices, (a) \textit{cross-sectional design},  (b) \textit{incident user design}, and (c) \textit{incident user design with active comparators}. At different points in time, a person either does not have a regular eating partner (marked in gray), or she does (marked in green, red, or blue). A cross-sectional design observes food consumption at the start of the monitored period, a fixed time $t_0$, which is the same across all participants. Incident user design isolates the effect of the onset of co-eating with another person on subsequent food consumption. In incident user design, time is tracked relative to the moment of onset $t_0$, which may be different across participants. The active comparator design additionally allows for comparisons of the effect of onset among persons who all start to eat with someone, but their partners have different characteristics (marked in red and blue). The present paper is based on an incident user design with active comparators (presented in more detail in Figure~\ref{fig:diagram2}).}
    \label{fig:1}
\end{figure}

Recall that we are interested in determining whether and how eating with others impacts the nature of food consumption. As depicted in Figure~\ref{fig:1}(a), a na\"ive approach to answering those questions would be a \textit{cross-sectional design}: at any given absolute point in time, some people are regularly eating with their peers (indicated with green) while others do not (indicated with gray). Starting from a certain absolute point in time $t_0$, by identifying persons with different habits, one could compare what is consumed by the persons who do not have a regular eating partner with what is consumed by the persons eating with a regular eating partner. One could also compare the food consumed by persons who are eating regularly with partners who have different habits.

The problem with this setup is that those persons who do not eat with others might have done so in the past (\eg, Person~1 in Figure~\ref{fig:1}(a)). Those who do eat with others might have been doing it for a long time or might have just initiated. Also, some people stop eating with others, whereas other people continue. It could be that those who stop do so because they prefer the diet they seek when eating alone (\ie, selection bias). Additionally, people who eat with others might differ in fundamental ways from those who do not. 

For these two reasons, looking at everyone at the same moment in a cross-sectional way can be problematic. To overcome these challenges, we can turn to an \textit{incident user design} (Figure~\ref{fig:1}(b)), which restricts the population to those people who newly initiate the treatment---starting to eat together with another person. We are interested in the causal effect on food consumption of initiating eating with a peer. Among people who had no regular eating partners in the past, what is the causal effect of starting to eat with a peer? In this way we isolate the causal effect of initiation. We restrict the observed population so that none of the persons have a history of eating with someone. Note how Person~1 in Figure~\ref{fig:1}(b) starts eating together with a regular partner, but then after a while no longer has a regular eating partner. This is not an issue because we are interested in the effect of the onset.

\begin{figure}[t]
        \includegraphics[width=0.8\textwidth]{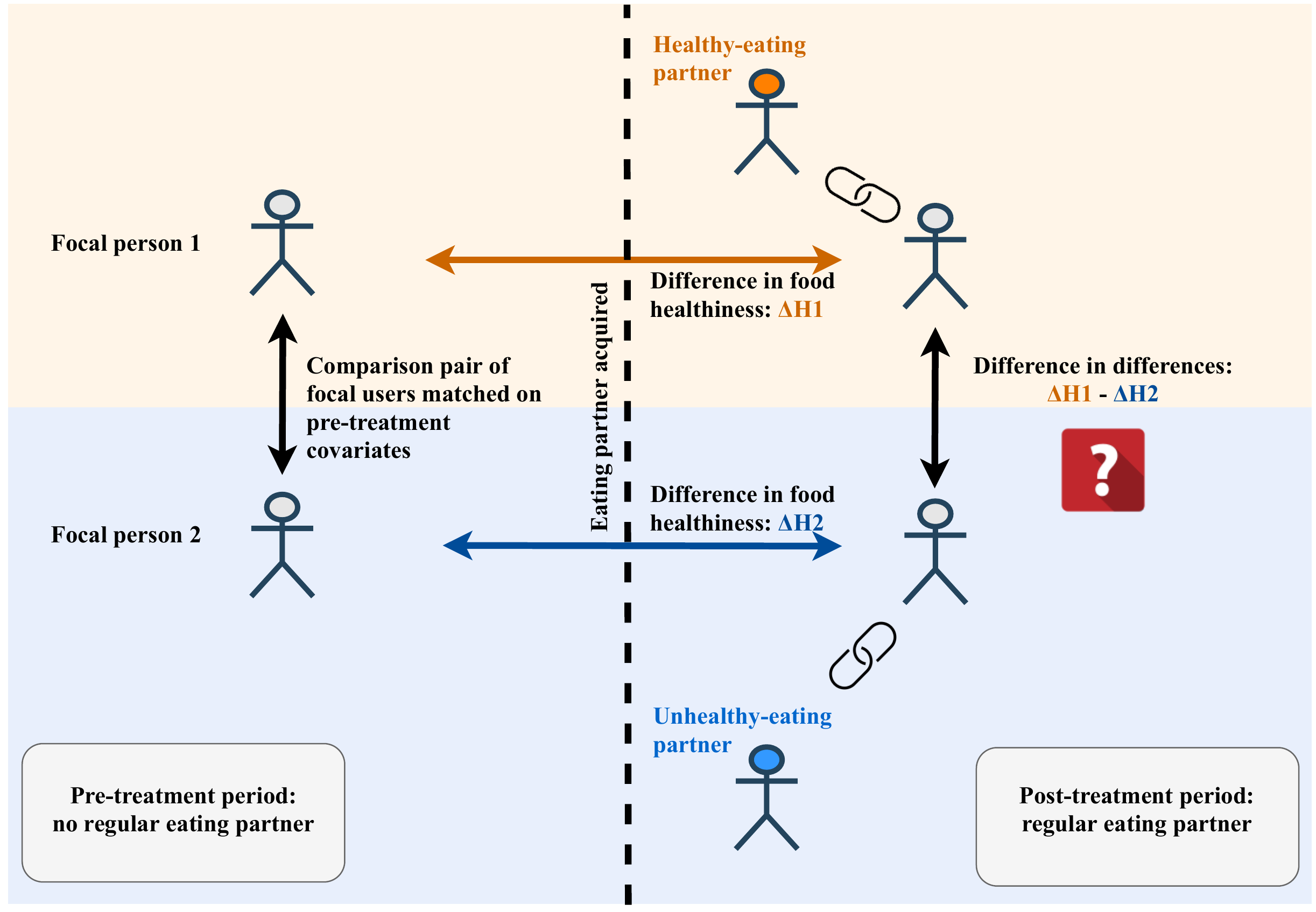}
    \caption{The matched incident user design with active comparators on which the present study is based. We identify comparison pairs of focal persons 1 and 2, who are indistinguishable in the pre-treatment period and have no regular eating partners, until the moment of co-eating onset, when they each acquire a regular eating partner. Focal person 1 starts regularly eating with a healthy-eating partner, while focal person 2 starts regularly eating with an unhealthy-eating partner. The comparison pair of focal users is then observed in the pre-treatment period (no regular eating partner) and post-treatment period (regular eating partner). The effect of the co-eating onset is estimated using a difference-in-differences analysis.}
    \label{fig:diagram2}
\end{figure}

As opposed to the cross-sectional design, where time is absolute, the incident user design offers the flexibility of tracking time relative to an onset $t_0$ that may be different for different participants. Although this design allows us to compare different treatments, the problem with this setup, which persists from the above-described cross-sectional design, is that, if the comparison group is ``no treatment'' (\ie, no initiation of co-eating), it is not apparent when the follow-up should start for the ``no treatment'' group. Additionally, selection bias remains and is not accounted for, as people who do not initiate might in other fundamental ways differ from those who do initiate.

Our study design addresses these challenges by implementing a variant of incident user design, \textit{incident user design with active comparators} (Figure~\ref{fig:1}(c)). Here, before initiation, no user included in the study had a regular eating partner (\ie, was treatment-free). We compare the effect of initiating to eat with partners who have different habits among persons who all initiate to eat with someone (illustrated with blue and red in Figure~\ref{fig:1}(c)). Active comparator designs tend to involve significantly less confounding \cite{yoshida2015active,lund2015active,johnson2013incident}, as people who eat with different kinds of others are more alike among themselves than when compared to people who do not have regular eating partners.

Our study design is illustrated in more detail in Figure~\ref{fig:diagram2}. We identify persons (referred to as \textit{focal persons}) who had no regular eating partners and, at a moment $t_0$ specific to that focal person, initiate eating with someone (referred to as \textit{eating partner}).
Here, as defined in Section~\ref{sec:Inference of co-eating onset from proximity in transaction logs}, a person qualifies as a focal person's potential eating partner if the two were observed making subsequent purchases in the same queue within one minute of one another on at least 10 occasions in the entire dataset,
and the onset of co-eating is defined as the first one of these occasions.
We then isolate pre-treatment and post-treatment periods of the focal person's food purchases comprising all transactions made six months before the first purchase together (moment $t_0$) and six months after, respectively. We ensure that the focal person does not initiate eating with anyone else in the pre- and post-treatment six months. The length of the pre-treatment period is chosen so that it is feasible to expect that an individual will be present on campus given the typical stay in the logs (the total observed 12 months of pre- and post-treatment correspond to one school year).

\begin{figure}[t]
    \includegraphics[width=0.7\textwidth]{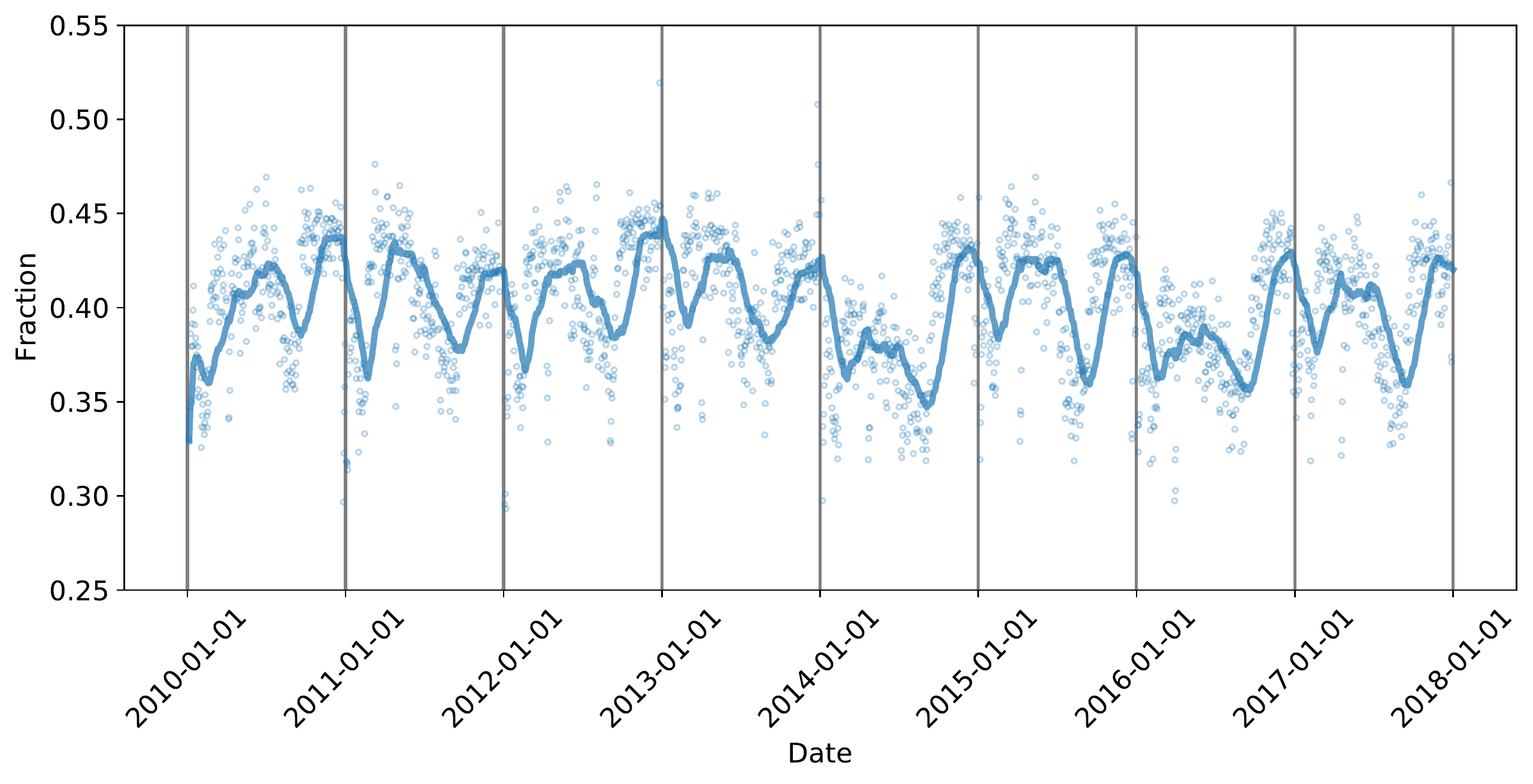}
    \caption{Daily fraction of purchases annotated as potentially unhealthy, tracked over five years. A seasonal pattern emerges. Drops in the daily fraction of unhealthy purchases coincide with between-semester breaks.}
    \label{fig:4}
\end{figure}

Some persons initiate co-eating with a person who has a positive healthiness score in the aligned pre-treatment period. In contrast, some initiate co-eating with a partner who has a negative score. These are the two groups that we seek to compare (we refer to the two types of partners as \textit{healthy-eating partner} and \textit{unhealthy-eating partner}).

For a focal person who starts to eat together with a partner who has a healthy dietary pattern, an \textit{active comparator} (or \textit{counterpart}), will be another focal person who starts to eat together with a partner who has an unhealthy dietary pattern. The potential counterparts start to eat with their partner in the same month as the other counterpart.
This is done in order to control for temporal confounds that might arise from a seasonal variation of food popularity: as seen in Figure~\ref{fig:4}, unhealthy foods are especially popular at certain times of the year.
The healthiness of the partner's dietary pattern is determined according to its numeric value (greater or less than zero), and not relative to the focal person.

Comparing incident users with active comparators is an important step towards reducing the impact of biases. However, in the assignment of the type of treatment, there can still be confounding. For example, it might be the case that only people who already have healthy habits start eating together with a partner who has healthy habits, due to a preference for similar others. The influence of the partners would then be indistinguishable from the impact of selection biases caused by homophily. 

Hence, we turn to a \textit{matched} incident user design with active comparators. We introduce an improvement over the previously discussed setup, where the incident users are matched to the potential active comparators while additionally controlling for pre-treatment covariates. Our goal here is to balance potential confounding variables within pairs, to be able to observe how the onset of co-eating with partners with different dieting patterns is associated with subsequent changes in the focal person's dieting pattern. We achieve this by performing a propensity-score-based causal analysis. We approximate randomized treatment assignment by modeling the propensity to experience the assigned intervention, relying on a number of pre-treatment covariates describing the focal persons' eating profiles. Due to the balancing property of propensity scores \cite{10.1093/biomet/70.1.41}, matching on propensities results in similar covariate distributions between groups that differ in their assigned interventions.

The covariates capture important dimensions of the pre-treatment dietary pattern of the focal person:
\textit{where} the food is purchased (what is the shop where the person most frequently buys food),
\textit{when} the food is purchased (what is the fraction of items occurring during lunchtime),
\textit{what} types of items are purchased (what fraction of purchased items are meals, and what is their estimated healthiness),
and \textit{how often} the person purchases food on campus (number of transactions).
We measure these confounding covariates up to time $t_0$.

\begin{figure}[t]
    \begin{minipage}{.45\textwidth}
    \centering
    \includegraphics[width=\textwidth]{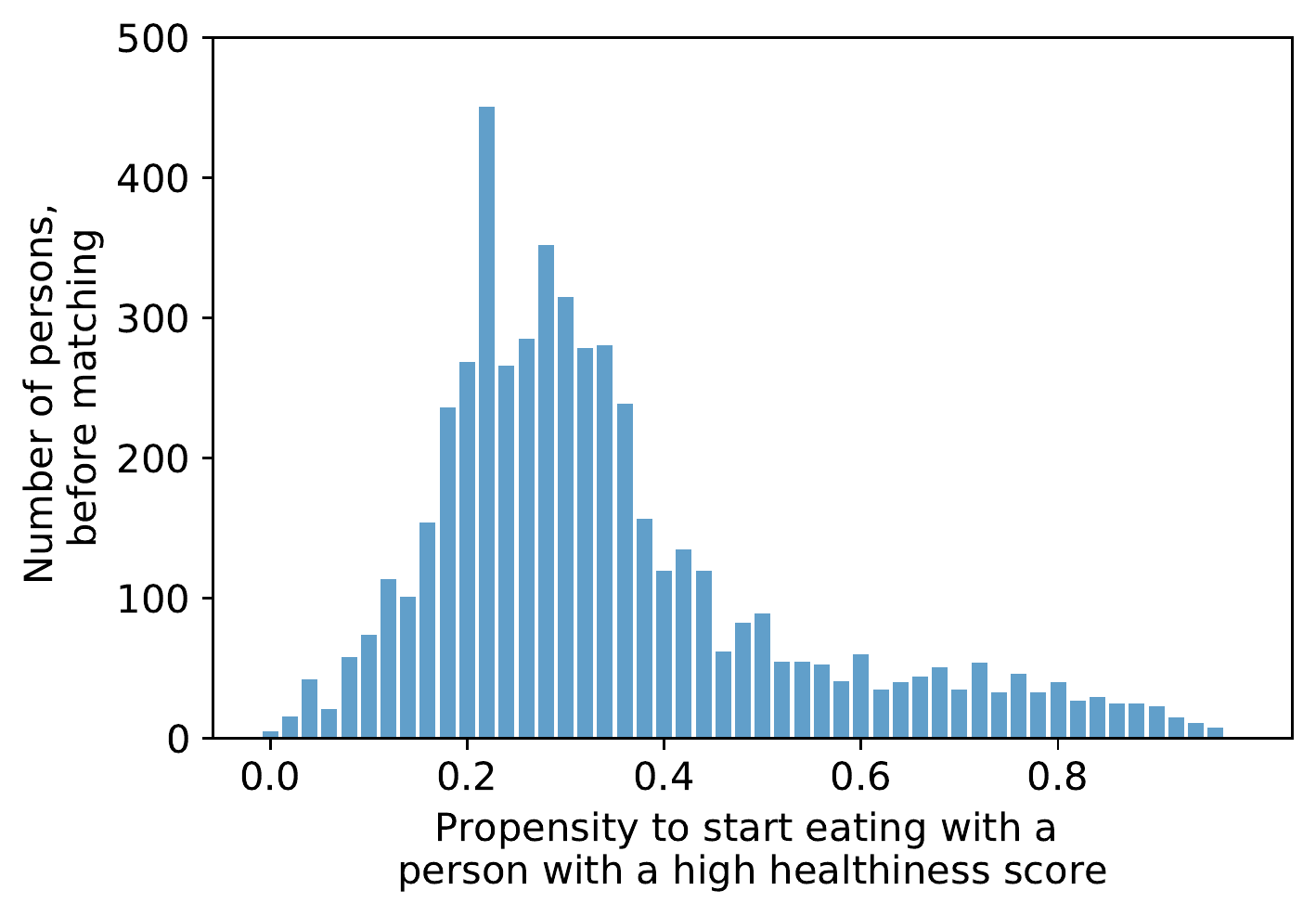}
    \caption{Distribution (before matching) of propensity to start eating with a healthy-eating partner.}
    \label{fig:7a}
    \end{minipage}
    \hspace{.05\textwidth}
    \begin{minipage}{0.45\textwidth}
    \centering
        \includegraphics[width=\textwidth]{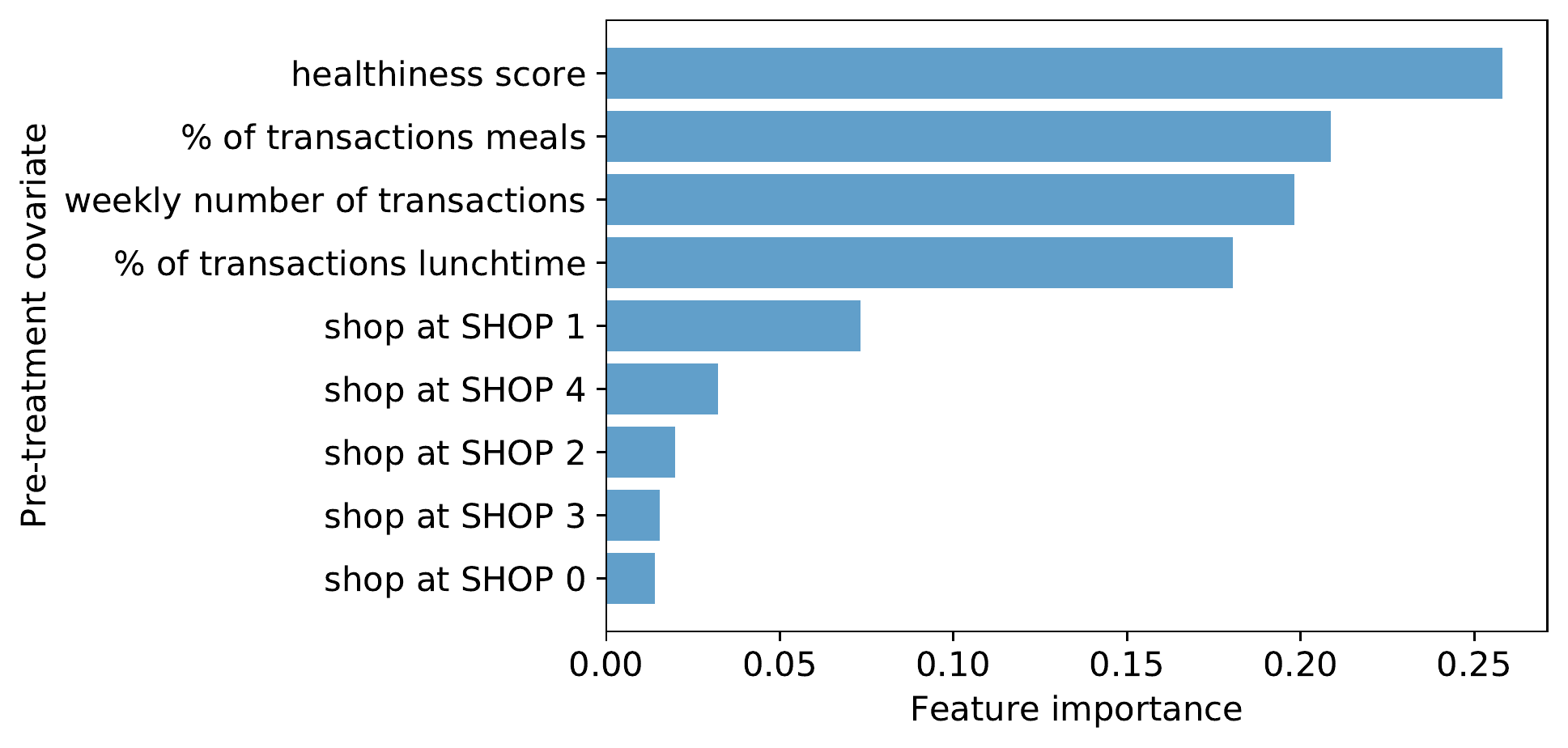}
        \caption{Importance of most indicative features for predicting treatment assignment, \ie, initiation of eating with a healthy- \vs\ unhealthy\hyp eating partner (shop names anonymized). Most important feature: pre-treatment healthiness score of focal person's purchases, which indicates homophily.}
        \label{fig:7b}
    \end{minipage}
\end{figure}

We use a random forest model that predicts the type of treatment based on pre-treatment covariates of the focal person (area under the ROC curve: 0.87). This implies that past purchases allow us to accurately predict whether the tie will be formed with a healthy- or an unhealthy-eating partner, and that confounding is a real problem that needs to be addressed.
The distribution of the propensity to start eating with a partner who has a high healthiness score is presented in Figure~\ref{fig:7a}. We also examine the feature importances in predicting the treatment assigned, \ie, the initiation of eating with a partner who has a healthy or unhealthy eating pattern (Figure~\ref{fig:7b}). We observe that the focal person's pre-treatment healthiness score is in fact the most important predictor of the type of partner the focal person will start to eat with, pointing at homophily.
	
Focal persons in the two sets are then matched while ensuring that two potential matches have propensity scores (likelihoods of receiving the treatment) within a caliper of 0.1. The size of the caliper was chosen so that balance in covariates is achieved. Moreover, an exact match on the sign of the mean pre-treatment healthiness score and the most frequented shop is required to achieve tight control. We then create matched pairs based on possible candidates by performing maximum weight matching on the weighted bipartite graph, where nodes are focal persons, and the weights use similarity based on the Mahalanobis distance in covariates. We maximize the total similarity to find a maximal matching.

\begin{figure}[t]
    \begin{minipage}{.45\textwidth}
\centering
\includegraphics[width=0.9\textwidth]{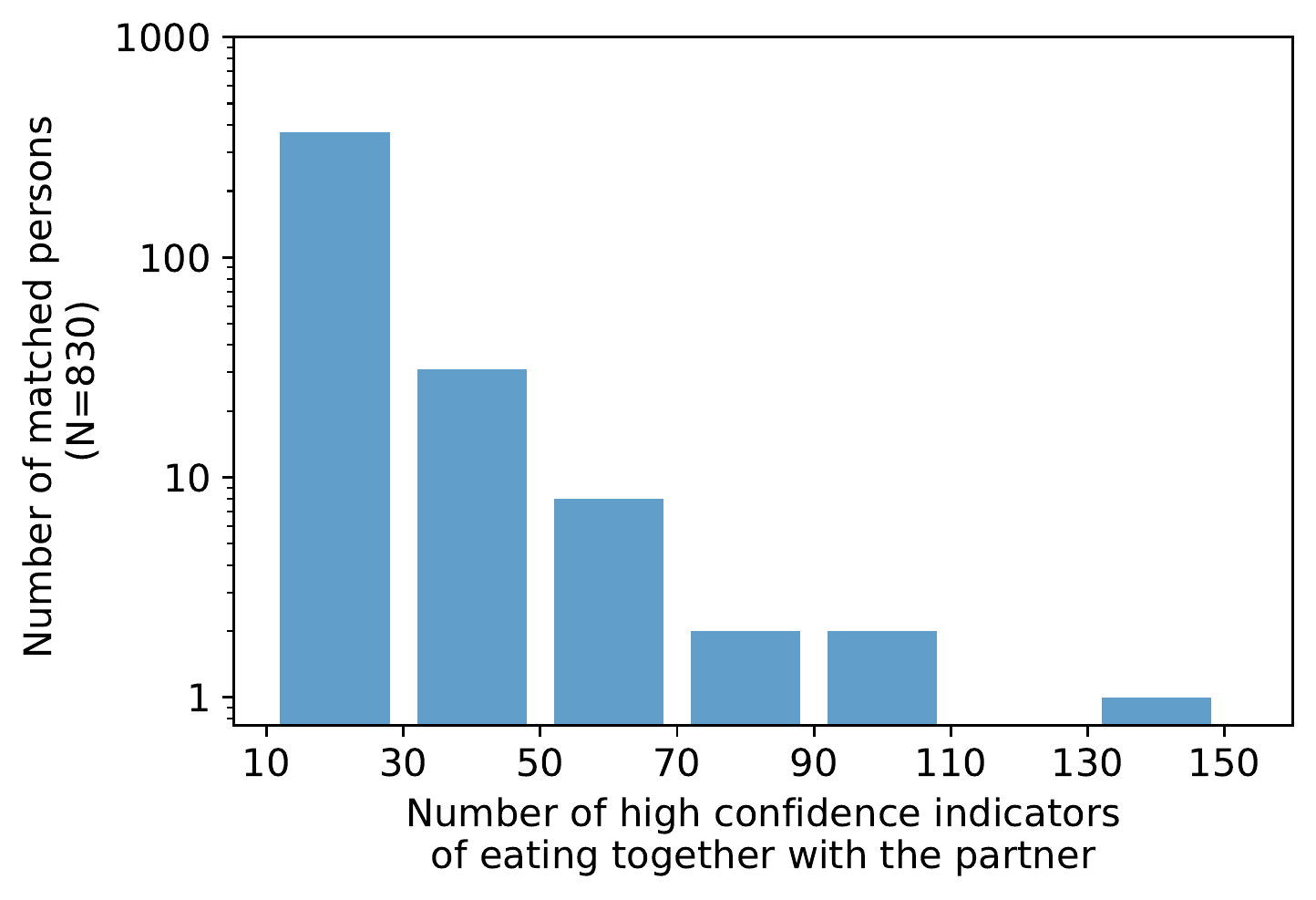}
        \caption{Histogram of the number of high confidence indicators of eating together with their respective partners, for matched focal persons.}
        \label{fig:5}
    \end{minipage}
    \hspace{.05\textwidth}
    \begin{minipage}{.45\textwidth}
\centering
        \includegraphics[width=0.9\textwidth]{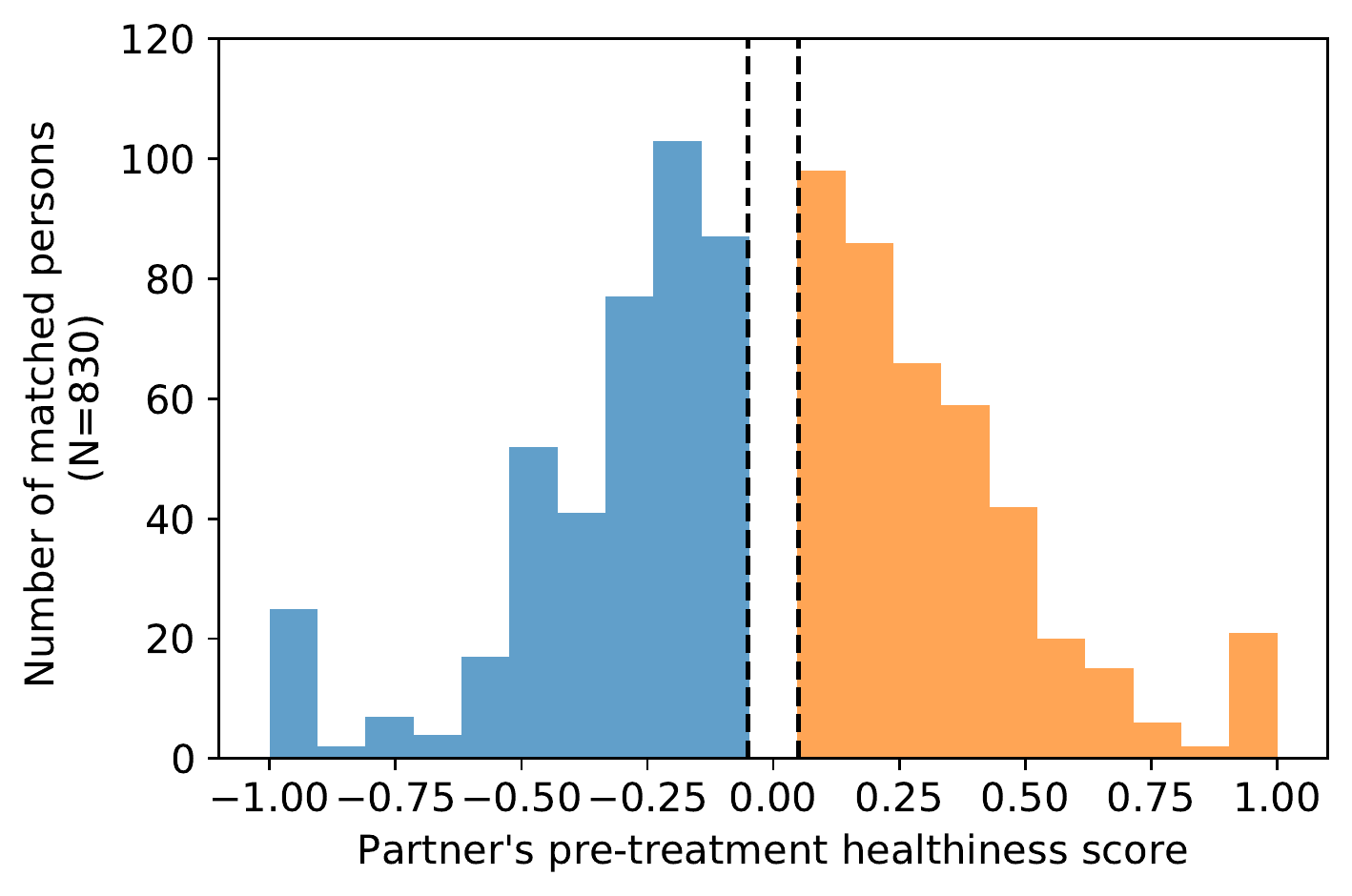}
        \caption{Histogram of the partner's pre-treatment healthiness score, across matched persons. Orange bars correspond to healthy-eating partners, blue bars to unhealthy-eating partners. A margin of 0.1 is ensured to differentiate the treatments.}
        \label{fig:6}
    \end{minipage}
\end{figure}

The result is a set of matched pairs of focal persons, indistinguishable up to the moment of initiation, who initiated co-eating with partners with different dietary patterns in the same month. This approach yielded 415 matched pairs of 830 focal persons who started to eat with different partners. We require at least 10 high confidence indicators of eating together with the partner (Figure~\ref{fig:5}). Partners' distribution of pre-treatment healthiness scores is shown in Figure~\ref{fig:6}.

Our matched analysis then moves on to comparing focal people who initiate co-eating with a person with a healthy dieting pattern, to their counterparts who have the same dieting patterns up to the moment of initiation, but initiate co-eating with a partner who has an unhealthy dieting pattern. The post-treatment patterns are then compared across treatments within the matched population.

\begin{table}[b]
    \small
  \caption{To ensure that matched persons are comparable, we evaluate the balance of their pre-treatment covariates, via the standardized mean difference (SMD) across covariates in the two matched groups.}
  \label{tab:1}
\begin{tabular}{l|c|c}
    \textbf{Pre-treatment covariate} & \textbf{SMD before matching} &  \textbf{SMD after matching} \\
    \hline
    Preferred shop (\ie, where the largest fraction of  &  \multicolumn{2}{c}{exact match required} \\
    pre-treatment transactions is made) & \multicolumn{2}{c}{}  \\
    \hline
    Pre-treatment percentage of lunchtime transactions & 0.109 & 0.045\\
    \hline
    Pre-treatment percentage of meal transactions & 0.207 & 0.075\\
     \hline
    Pre-treatment mean healthiness score & 0.301 & 0.023\\
    \hline
    Pre-treatment mean weekly number of transactions & 0.071 & 0.023 \\
  \end{tabular}
\end{table}

Before moving on to the analysis of the outcomes, we ensure that the matched persons are comparable by measuring the balance of their pre-treatment covariates (Table~\ref{tab:1}). We use the standardized mean difference (SMD) across covariates in the two groups to measure the balance. We observe that matching greatly reduces the SMD, as the largest SMD across covariates (the one of the pre-treatment healthiness score of the focal person) changes from 0.301 before matching, to 0.023 after matching. Groups are considered balanced if all covariates have SMD lower than 0.2~\cite{kiciman2018using}, a criterion that is satisfied here.

\section{Results}
\label{sec:results}


Recall our research question: we want to understand how a person's food choices are affected by the healthiness of a co-eating partner's food choices. Do people's choices change and, if so, in what direction (\ie, towards more or less healthy)?


\subsection{Regression analysis of pooled data}

\begin{figure}[t]
    \begin{minipage}{\textwidth}
    \centering
    \includegraphics[width = 0.8\textwidth]{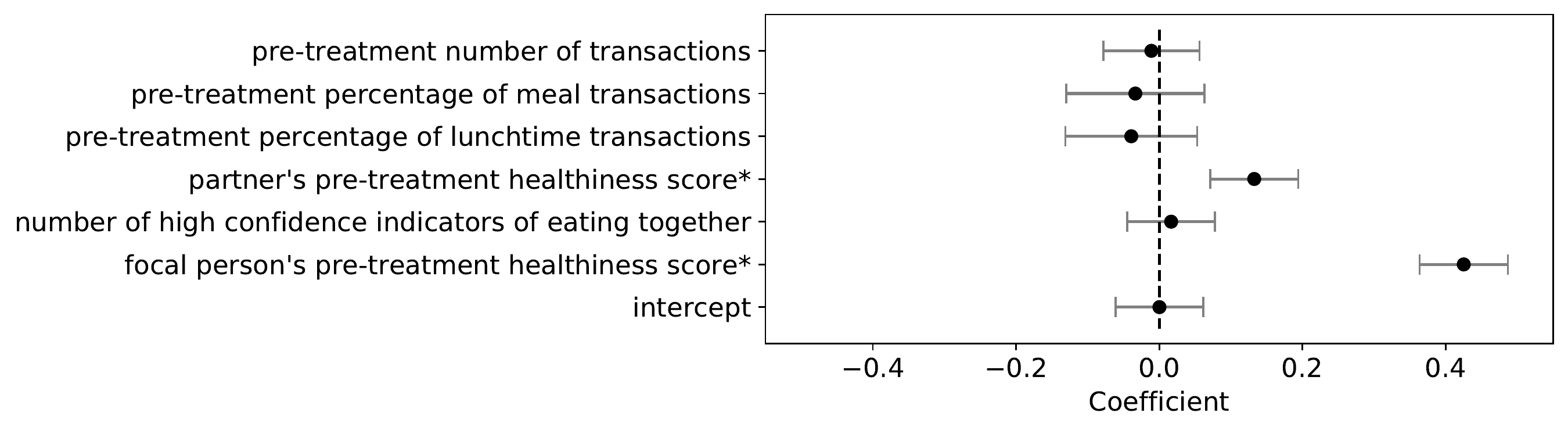}
    \caption{The effect of the focal person's pre-treatment covariates, the partner's pre-treatment healthiness score, and the number of detected high-confidence indicators of eating together on the focal person's post-treatment healthiness of purchased items. The effects are estimated with linear regression ($R^{2}=0.194$); 95\% confidence intervals are approximated as two standard errors. Significant coefficients ($p<0.05$) are marked with an asterisk (*). The focal person's own healthiness score and their eating partner's healthiness score are the only two statistically significant factors associated with the focal person's post-treatment healthiness score.}
    \label{fig:8}
     \end{minipage}
\end{figure}

First, we aim to determine if there are any significant differences between the outcomes of the matched focal persons. Is the pre-treatment healthiness of the partner predictive of post-treatment healthiness of the focal person? We start by performing a regression estimation of the effect of the partner's pre-treatment healthiness score on the focal person's post-treatment score.

We fit a model where the focal person's post-treatment healthiness score is the dependent variable. We include the focal person's pre-treatment healthiness score, the partner's pre-treatment healthiness score, the number of high-confidence indicators of eating together, and the focal person's pre-treatment covariates as the independent variables. The focal person's pre-treatment covariates (number of transactions, percentage of transactions that are meals, percentage of lunchtime transactions, and the pre-treatment healthiness score) are already controlled for by matching, but they are included in the model to account for possible residual confounding. The predictors and the outcome are standardized, so the coefficients are interpreted as increases in the healthiness score per standard deviation of the predictor.

In Figure~\ref{fig:8}, fitting the linear regression, we measure a significant positive effect of 0.13 (95\% CI [0.07, 0.19]) of the partner's pre-treatment healthiness score. The focal person's own pre-treatment healthiness is the strongest predictor of post-treatment healthiness (coefficient 0.43, 95\% CI [0.36, 0.49]). This is the first indicator that the pre-treatment score of the partner is associated with the focal person's patterns.


\subsection{Contingency-table analysis}

Next, to obtain fine-grained insights about patterns taking place at the pair level, we analyze the outcomes with a contingency table.
We binarize the outcome to look for either an increase or no increase post-treatment, compared to pre-treatment. Four possibilities exist for any matched pair: both increased, only one or the other increased, and none increased. The contingency table is presented as Table~\ref{tab:2}. The table counts the frequency of the four possible results. Using a chi-squared test, we reject the null hypothesis of no treatment effect ($p = 0.00017$).

It is particularly informative to observe the discordant pairs (off-diagonal entries in the contingency table) among the matched pairs. Such pairs correspond to situations when the outcome (increase or no increase) differs in the matched pair. The intuition is the following: if there is no effect, the two types of discordant entries should be balanced. However, we observe that in 103 pairs, the focal person with a positive intervention increased, and the focal person with a negative intervention did not. The reversed situation, in comparison, occurs in 67 pairs. We test the null hypothesis of no effect in a paired randomized experiment using McNemar's test \cite{lachenbruch1998assessing}, which relies directly on the evidence that comes from the discordant pairs (their number and the ratio between them). Here, too, we reject the null hypothesis of no treatment effect ($p = 0.007$).

\begin{table}[t]
    \small
  \caption{Contingency table counting number of pairs of matched focal persons in each condition. Post-treatment healthiness score is compared to pre-treatment score to determine if there was an increase; in columns, for focal persons who start to eat with healthy-eating partners, and in rows, for their matched counterparts, \ie, focal persons who start to eat with unhealthy-eating partners.}
  \label{tab:2}
\begin{tabular}{l r c c | c}
    \multicolumn{2}{c}{} & \multicolumn{2}{l|}{\textit{Focal person with a }} & \\
    \multicolumn{2}{c}{} & \multicolumn{2}{l|}{\textit{healthy-eating partner}} & \\
    \multicolumn{2}{c}{} & \multicolumn{1}{c}{\textbf{Increase} } & \textbf{No increase} &  \textbf{Total pairs} \\
    \textit{Focal person with an   } &   \textbf{Increase} & 126 &  67 & 193 \\
    \textit{unhealthy-eating partner} &   \textbf{No increase} & 103 & 119 & 222 \\
    \hline
   \multicolumn{2}{r}{\textbf{Total pairs}}  & 229&  186& \textbf{415} \\
  \end{tabular}
\end{table}


\subsection{Difference-in-differences analysis}

We move on to further exploit the matched setup in order to estimate the  difference\hyp in\hyp differences~\cite{lechner2011estimation} effect for pairs of matched focal persons.
The idea is to first calculate the difference between post-treatment and pre-treatment healthiness scores for each focal person separately.
Then, we can calculate the difference in treatment effects between two matched focal persons in each pair.
Averaging the differences in differences across all pairs yields the overall treatment effect.

\xhdr{Regression model}
In practice, following the standard approach, we estimate the difference-in-differences effect via a regression model.
Here, each focal user adds two data points (one pre-treatment, one post-treatment), each of which specifies, as predictors, the type of partner with whom the focal user started to eat as a treatment (healthy-eating or unhealthy-eating) and the time period (pre- or post-treatment); and, as the outcome, the healthiness score of the focal user's food choice during the respective period.
Each matched pair thus contributes four data points, and the modeled dataset consists of $4 \cdot 415 = 1{,}660$ data points.
Formally, the model takes the following form:
\begin{equation}
    y_{it} = \alpha + \beta \cdot \text{healthy\_treatment}_i + \gamma \cdot \text{treated}_t + \delta \cdot (\text{healthy\_treatment}_i \cdot \text{treated}_t) + \text{error}_{it},
    \label{eqn:formula_overall}
\end{equation}
where the dependent variable $y_{it}$ is the focal user $i$'s healthiness score in period $t$,
and the independent variables indicate whether $i$'s partner has a positive or negative pre-treatment healthiness score ($\text{healthy\_treatment}_i$ = 1 or 0, respectively)
and whether the respective data point captures the pre- or post-treatment period ($\text{treated}_t$ = 1 or 0, respectively).
The coefficient $\delta$ of the interaction term, then, is the difference\hyp in\hyp differences effect of starting to eat with a healthy- \vs\ unhealthy\hyp eating partner.

\xhdr{Results}
Calculating the average difference-in-differences effect with a linear regression across all matched focal persons, we observe a larger post-treatment increase in focal persons with healthy-eating partners compared to the post-treatment increase in matched counterparts, $\delta=0.051$ (95\%~CI $[0.021,0.076]$, $R^2=0.07$). This means that, accounting for possible temporal drifts between post-treatment and pre-treatment that are not associated with the initiation, focal persons starting to eat with a healthy-eating partner significantly diverge from their matched counterparts starting to eat with an unhealthy-eating partner.
Quantitatively, the effect size of $\delta=0.051$ means that, compared to matched counterparts who start eating with an unhealthy\hyp eating partner, focal persons who start eating with a healthy\hyp eating partner increase their healthiness score by an additional 5.1\% of the full range spanning from a neutral healthiness score~(\ie, 0) to the maximum healthiness score~(\ie,~1).
           
Similarly, to estimate the effect of social tie formation on the absolute number of healthy and of unhealthy purchased items, we repeat the regression analysis described in \Eqnref{eqn:formula_overall},
but now with different dependent variables that capture the \textit{total number of healthy and the total number of unhealthy items purchased} by focal user $i$ in period $t$. We observe that the focal persons who start to eat with a partner with a healthy pattern purchase an additional 5.71 (95\% CI $[3.21, 8.21]$, $R^2=0.17$) healthy items, and an additional $-1.13$ (95\% CI $[-3.04, 0.78]$, $R^2=0.12$) unhealthy items in the six months following the tie formation, compared to their matched counterparts.


\xhdr{Sensitivity analysis}
The above finding relies on the assumption that there are no unobserved variables that create the differences between the matched focal persons that could explain away the measured effect.
Sensitivity analysis is a way of quantifying how the results of our calculations would change if the assumptions were violated to a limited extent.
If the conclusions of the study would change little, then the study is insensitive to a violation of the assumptions, up to the specified limited extent.
In contrast, if the conclusions would change substantially, then the study is sensitive to a violation of the assumption.

The key assumption made in our analysis is that the treatment assignment is not biased, or in other words, that after balancing the pre-treatment covariates, the co-eating initiation with a healthy-eating \vs\ an unhealthy-eating partner is randomized (\ie, it is effectively decided by a coin flip). We measure by how much that assumption needs to be violated in order to alter our conclusion that there is a significant difference\hyp in\hyp  differences effect on the healthiness of purchased items among the matched focal persons.
Specifically, sensitivity analysis lets us answer the following question: if there is a violation of randomized treatment assignment (\ie, a deviation from fair 50/50 coin flip), how large would it need to be in order to alter the conclusion that the null hypothesis of no difference between focal persons can be rejected?
This notion is quantified by the \textit{sensitivity}~$\Gamma$, which specifies the ratio by which the treatment odds of two matched persons would need to differ in order to result in a $p$-value above the significance threshold.
We always have $\Gamma \geq 1$, with larger values of $\Gamma$ corresponding to more robust conclusions.

For the chosen $p =0.05$, we measure a sensitivity of $\Gamma = 1.17$,
which implies that, within matched pairs,
an individual's probability of being the treated one could take on any value between
$1/(1+\Gamma) = 0.46$
and
$\Gamma/(1+\Gamma) = 0.54$
without changing our decision of rejecting the null hypothesis of no effect.
In other words, if the assignment of the treatment after matching were not approximating the ideal 0.5, but a third variable made some people more likely to initiate eating with a healthy-eating or an unhealthy-eating partner,
the randomized treatment assignment would have to be violated by deviating from the fair~0.5 by at least four percentage points.

\begin{figure}[t]
  \begin{center}
    \includegraphics[width=0.4\textwidth]{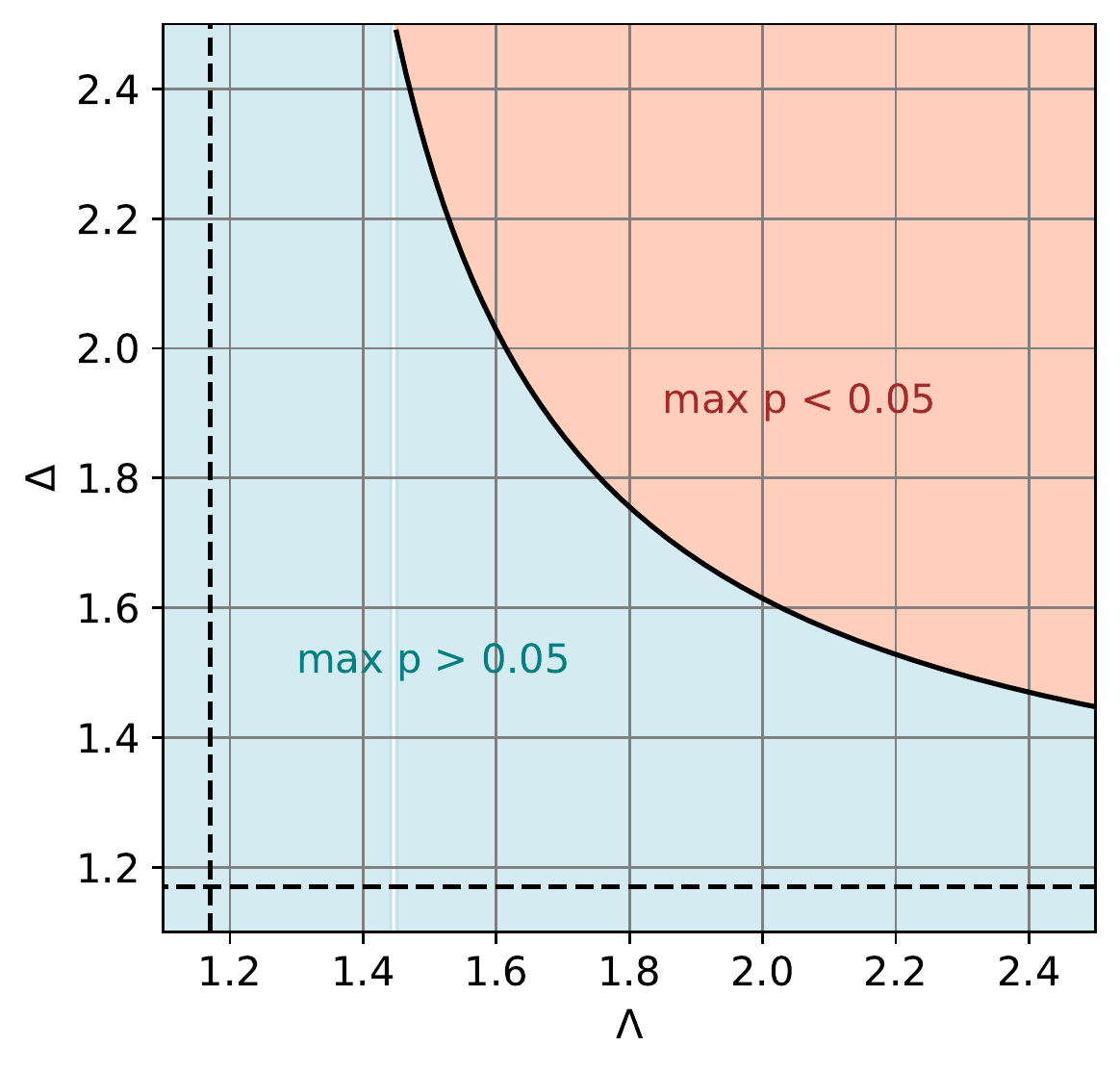}
    \vspace{-2mm}
    \caption{Sensitivity analysis. For the sensitivity $\Gamma = 1.17$, the amplification $(\Lambda,\Delta)$ is plotted (see text for explanation). Horizontal and vertical dashed lines indicate $\Gamma$, i.e., the asymptotic value of $\Lambda$ for $\Delta \rightarrow \infty$, and vice versa.}
    \label{fig:13}
  \end{center}
\end{figure}

Additionally, we conduct an amplification of the sensitivity analysis \cite{rosenbaum2009amplification}. Amplification is particularly relevant when the concern is not about a violation of the randomized treatment assignment, but rather about the potential existence of a specific unobserved covariate. It then becomes useful to consider possible combinations of $\Lambda$ and $\Delta$, two parameters describing the unobserved covariate, that would result in the measured $\Gamma$. The strength of the relationship between the unobserved covariate and the difference in outcomes within the matched pair is defined by $\Delta$, whereas $\Lambda$ defines the strength of the relationship between the unobserved covariate and the difference in probability of being assigned a treatment.
With these definitions, the 
sensitivity $\Gamma$ can be expressed in terms of $\Lambda$ and $\Delta$, as $\Gamma = (\Lambda \Delta + 1) / (\Lambda + \Delta)$.

The result of sensitivity analysis amplification is presented in Figure~\ref{fig:13}. For combinations of $\Lambda $ and $\Delta$ in the orange area, significant effects would be detected (leading to $p <0.05$), whereas for the combinations in the blue area, no significant effects would be detected (leading to $p >0.05$). An infinite number of $(\Lambda,\Delta)$ combinations fall on the border; \eg, $(\Lambda, \Delta) = (2.0,1.6)$ corresponds to an unobserved covariate that doubles the odds of treatment and multiplies the odds of a positive pair difference in the outcomes by 1.6.

Overall, we conclude that the study design is insensitive to small biases \cite{rosenbaum2017observation}.


\begin{figure}[t]
    \begin{minipage}[t]{.45\textwidth}
        \centering
    \includegraphics[width = \textwidth]{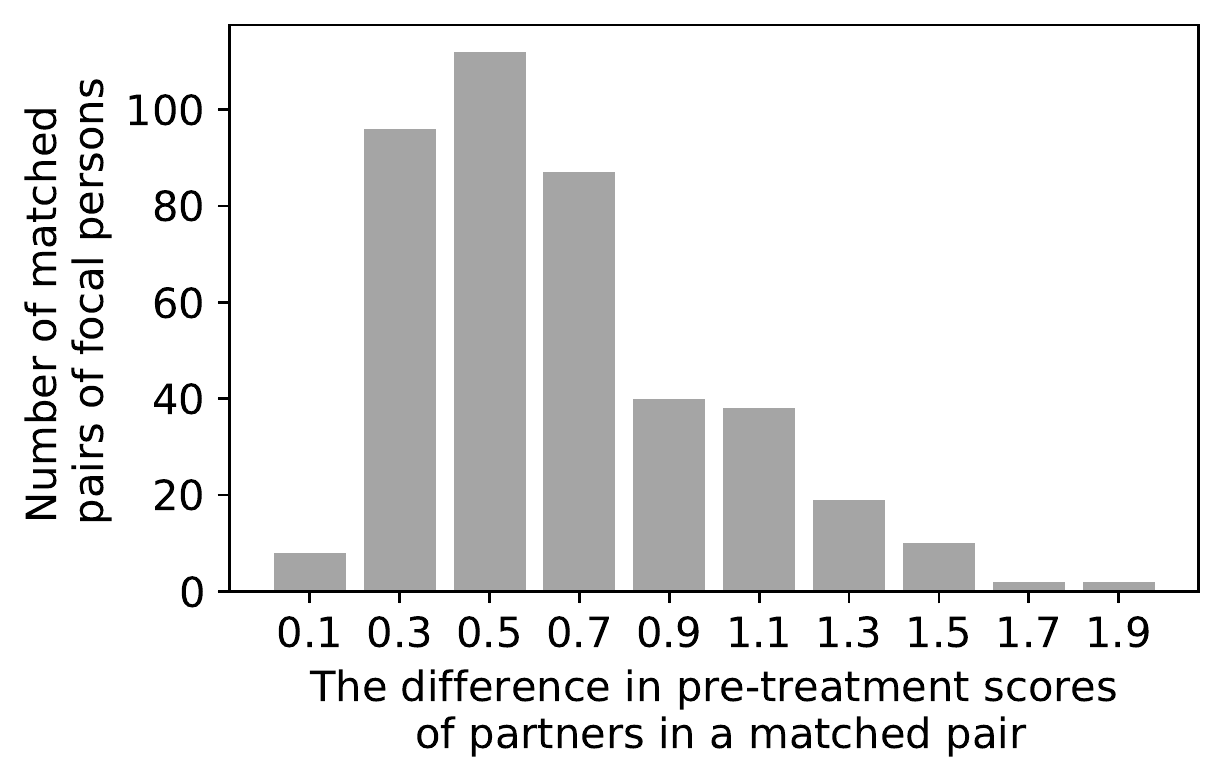}
        \subcaption{}\label{fig:dosea}
    \end{minipage}
    \hspace{0.04\textwidth}
    \begin{minipage}[t]{.45\textwidth}
        \centering
    \includegraphics[width =\textwidth]{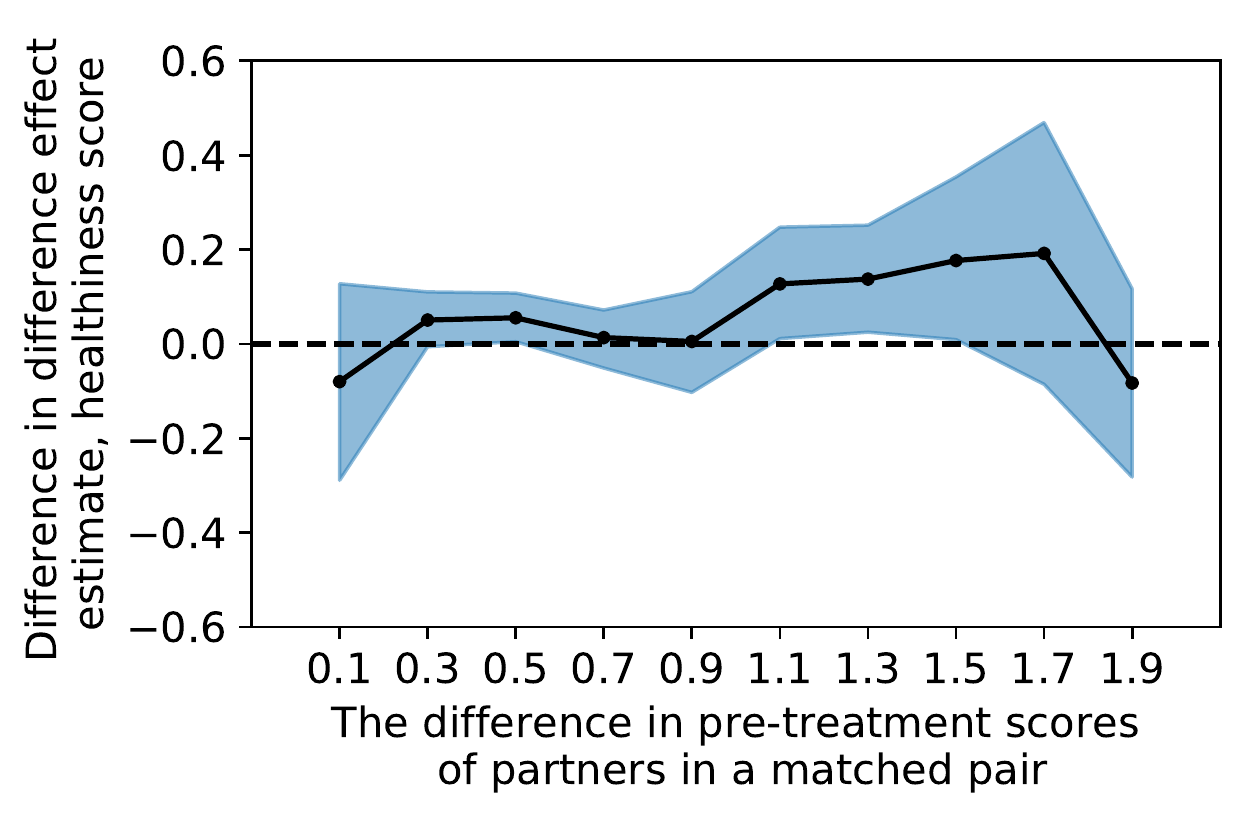}
        \subcaption{}\label{fig:doseb}
    \end{minipage}
    \hfill
    \vspace{-3mm}
    \caption{
Dose--response relationship.
\textbf{(a)}~Histogram of pre-treatment differences in healthiness scores between partners of paired focal users.
\textbf{(b)}~Difference\hyp in\hyp differences effect between focal persons in matched pairs, stratified by the pre-treatment differences in healthiness scores between the partners they were exposed to.
}
\end{figure}

\subsection{Dose--response relationship}

Next, we analyze the dose--response relationship in our matched setup. Similar focal persons initiate eating with differing partners. We observe systematic changes in the dieting patterns of the focal persons after the tie formation. But do more drastic difference\hyp in\hyp differences effects occur when the differences between partners are more drastic?
In the case of a true causal effect, one would expect a dose--response effect where focal persons diverge more post-treatment if their partners diverged more pre-treatment.

Although large differences in the pre-treatment scores between matched focal persons' partners are rare (Figure~\ref{fig:dosea}), Figure~\ref{fig:doseb} shows evidence of a dose--response relationship:
the difference\hyp in\hyp differences effect is stronger when the partners are more different (\ie, the more extreme difference in partners leads to more extreme effect estimates). If there were other confounding factors that could explain the observed difference\hyp in\hyp differences effects, and those factors had nothing to do with the onset of eating together, we would not expect to find a dose--response relationship.
The observed dose--response relationship thus further supports the conclusion of a causal effect.


\begin{figure}[b]
    \begin{minipage}{.32\textwidth}
        \centering
        \includegraphics[width=\textwidth]{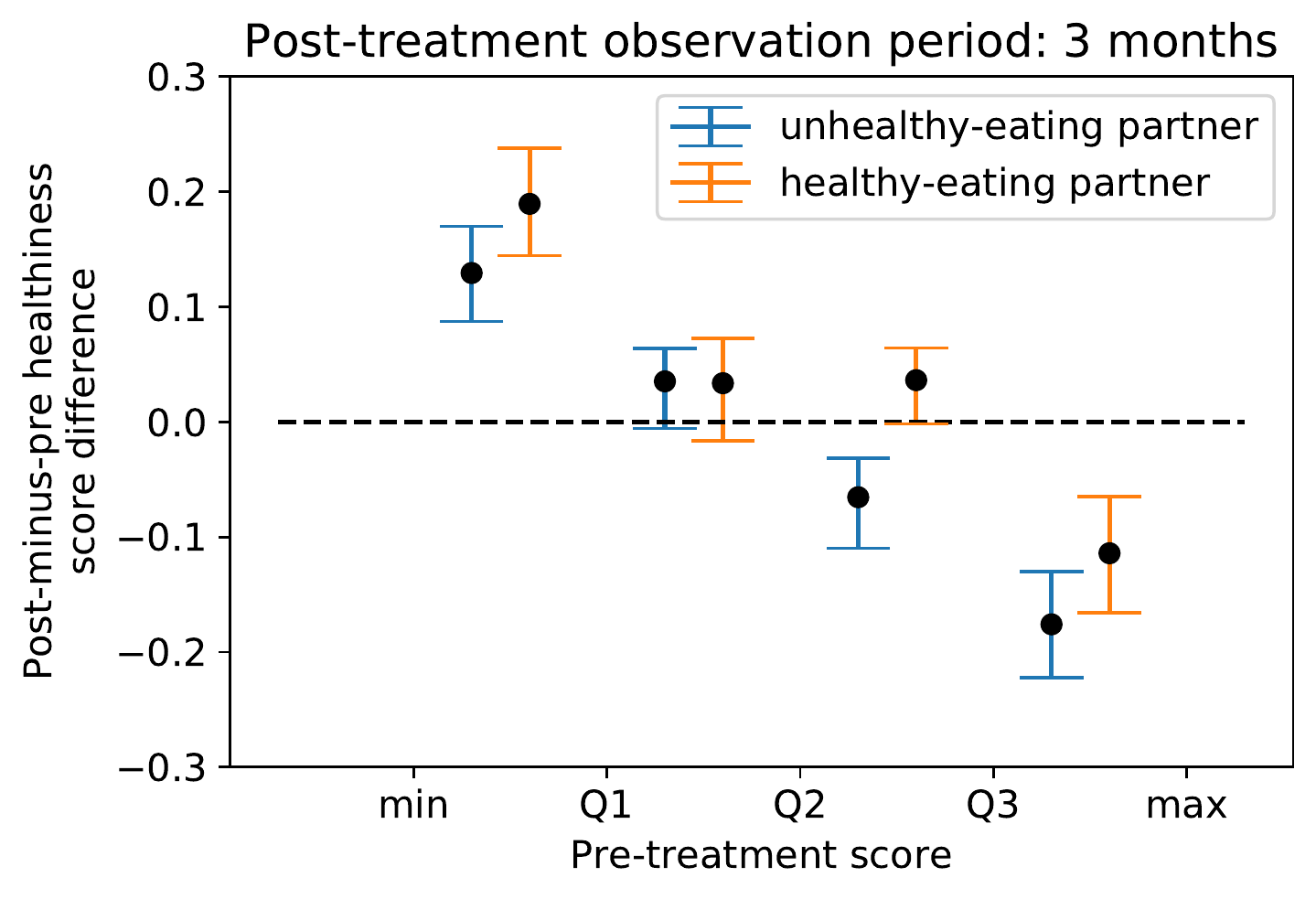}
    \end{minipage}
    \begin{minipage}{.32\textwidth}
        \centering
        \includegraphics[width=\textwidth]{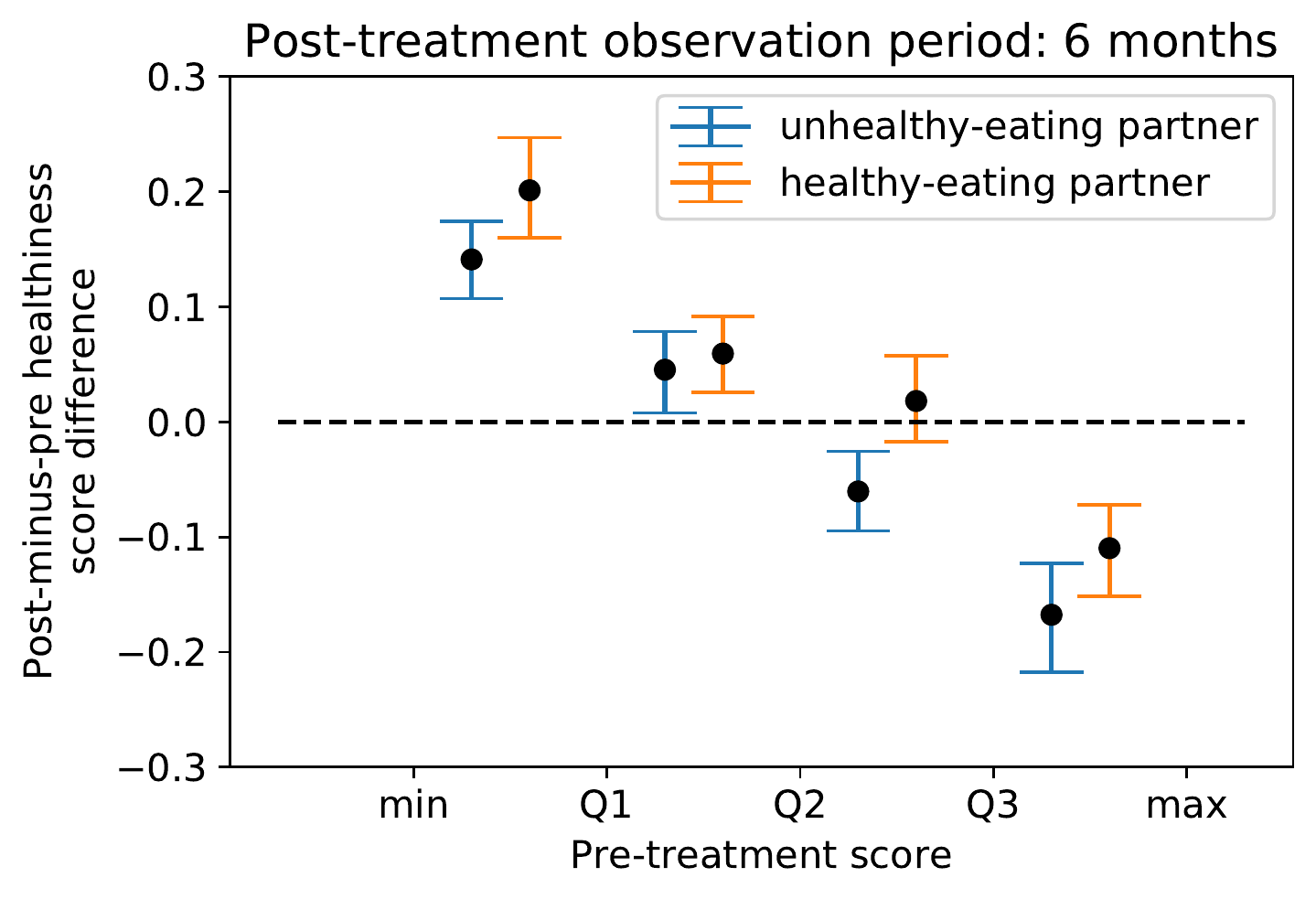}
    \end{minipage}
    \begin{minipage}{.32\textwidth}
        \centering
        \includegraphics[width=\textwidth]{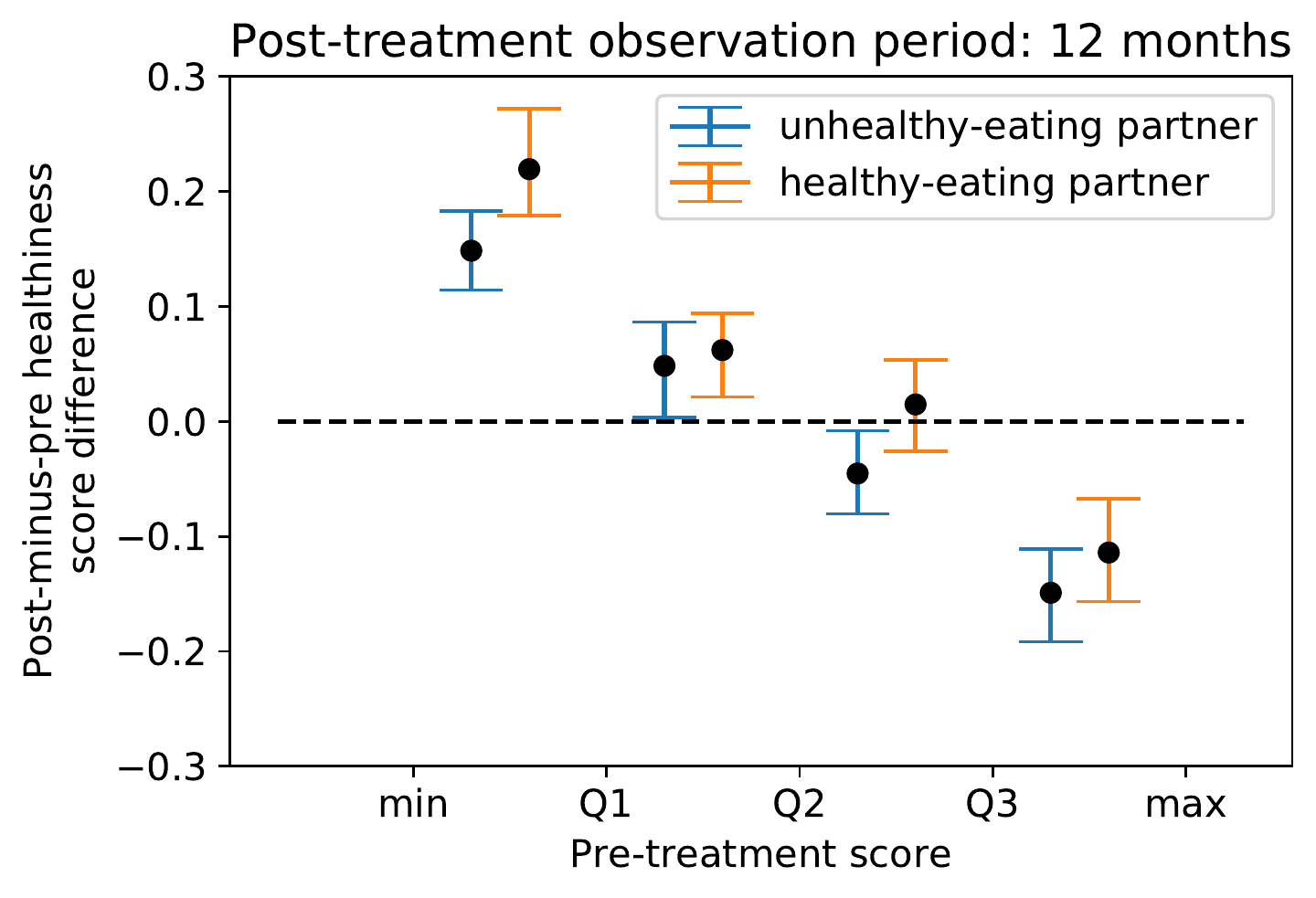}
    \end{minipage}
    \caption{\new{Post-treatment increase in healthiness score, stratified by pre-treatment healthiness score of focal person (with 95\% bootstrapped confidence intervals).} The difference is shown separately for focal persons who start eating with a person with a positive (orange)  \vs\ negative (blue) healthiness score. The difference is monitored in the first 3, 6, and 12 post-treatment months (left, center, and right panel, respectively).}
    \label{fig:10}
\end{figure}

\subsection{Stratification by pre-treatment healthiness}

Additionally, we would like to understand for whom the treatment is effective. Are there changes across the board with respect to the initial healthiness, or only for specific sub-populations? For whom is the intervention most efficient? We again monitor the differences between post- and pre-treatment healthiness scores, but now stratified into quartiles by pre-treatment healthiness score of the focal person (Figure~\ref{fig:10}).
\new{Moreover, we repeat this analysis for post-treatment observation periods of varying length (3, 6, and 12 post-treatment months).} In the aligned, post-intervention period, persons who start eating with partners with healthy dieting patterns are characterized with consistently higher healthiness scores compared to the matched counterparts, across strata of the focal person's pre-treatment healthiness score. Note that the fact that the slopes are decreasing may be a simple regression to the mean. \new{The key observation is that, within each stratum, when comparing the outcomes in orange and blue, people who initiate eating with a healthy-eating partner (orange) see a greater post- \vs\ pre-treatment difference compared to people who initiate eating with an unhealthy-eating partner (blue).}


\subsection{Analysis of affected food-item categories}

\begin{figure}[t]
    \begin{minipage}{\textwidth}
        \centering
        \includegraphics[width=0.7\textwidth]{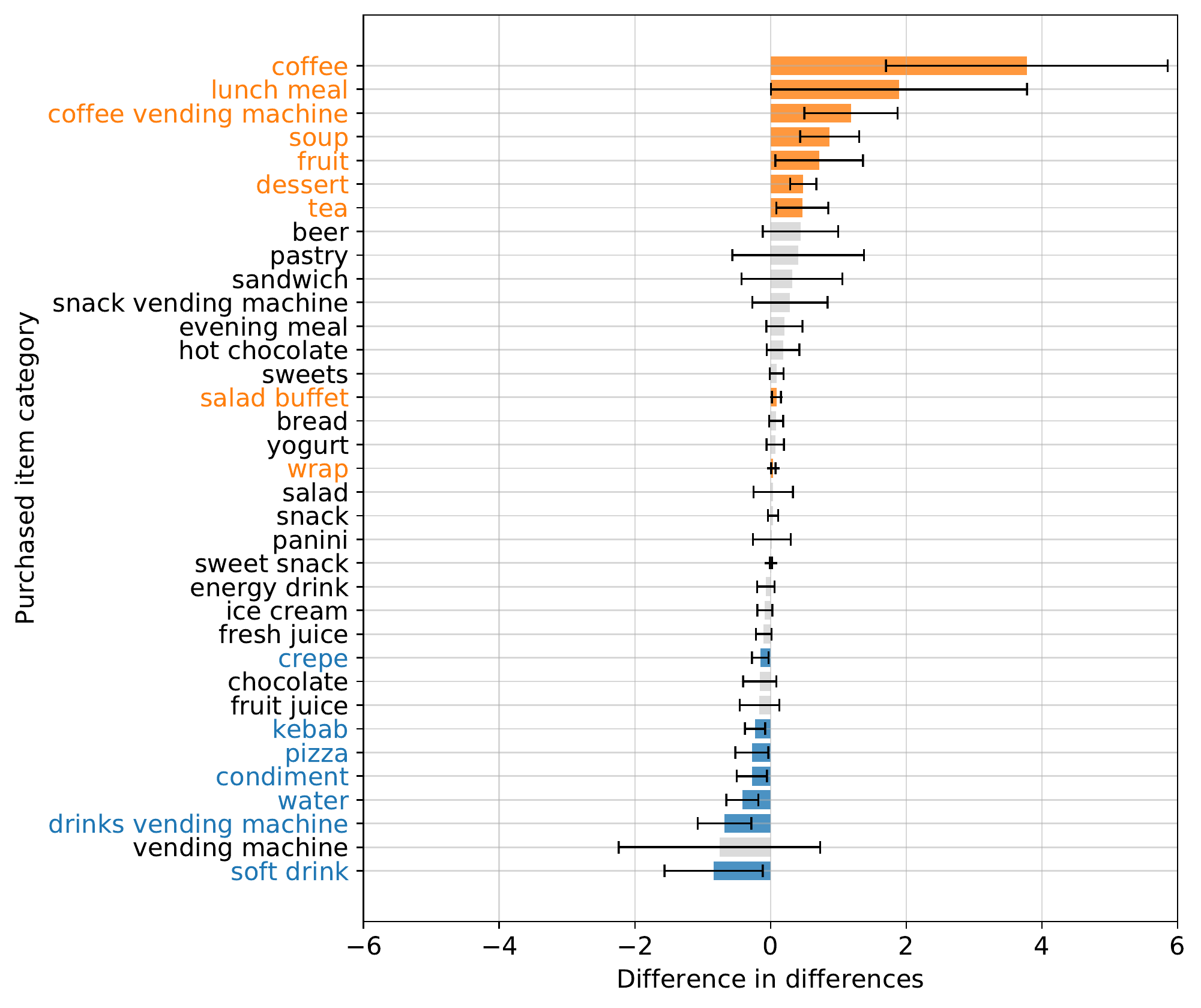}
        \caption{Estimated difference\hyp in\hyp differences effects of co-eating onset with healthy- \vs\ unhealthy-eating partner on frequency of purchased food categories (with 95\% confidence intervals approximated as plus\slash minus two standard errors). Categories with a significant effect are marked in orange (positive) and blue (negative), whereas categories with no significant effect are marked in gray.}
        \label{fig:12}
    \end{minipage}
\end{figure}

Finally, we set out to understand the influence of new co-eating partners on the rates at which categories of food items are subsequently purchased. Since we observed that the behaviors are modified, we now ask: what items are eaten more, and which less? What foods being purchased and eaten in group settings on campus have the largest influence on others?

To estimate category\hyp specific difference\hyp in\hyp differences effects, we repeat the regression analysis described in \Eqnref{eqn:formula_overall}, but now with a different dependent variable $y_{cit}$, which captures the number of items from food category $c$ purchased by focal user $i$ in period $t$.
By fitting a separate regression for each food category $c$, we obtain category-specific effects $\delta_c$.

The estimated effects $\delta_c$, together with 95\% confidence intervals, are presented in Figure~\ref{fig:12}. We observe that the focal persons initiating to eat with a healthy-eating partner purchase more coffee, lunch meals, coffee from vending machines, soup, fruit, dessert, tea, salad, and wraps, compared to their matched counterparts, who purchase more soft drinks, drinks from vending machines, water, condiments, pizza, kebabs, and cr\^epes. The values on the $x$-axis can be interpreted as the number of purchased items by which the matched focal persons diverge in the post-treatment period. For example, in the six months following tie formation, people who start eating with healthy-eating partners purchase, on average, around two additional meals and around four additional coffees, compared to the matched counterparts. The matched counterparts who start eating with unhealthy-eating partners, by contrast, on average purchase around one additional soft drink in the six months following tie formation.

Coffees and lunch meals are the items that see the largest increase after tie formation with a healthy-eating partner. These items are in general purchased in large numbers (Figure~\ref{fig:2}). Conversely, items with the strongest effect among the matched counterparts, with the exception of water, loosely form a cluster of potentially unhealthy items that should not be eaten in large quantities. The remaining items with a significant positive effect, soups, fruits, desserts, tea, salad buffet, and wraps are overall less indicative of an unhealthy dietary pattern.

\section{Discussion and conclusions}
\label{sec:discussion}
We report on a longitudinal observational study of the effect of the formation of social ties on food choice, leveraging a novel source of data: logs of millions of food purchases made over an eight-year period on a major university campus.
To estimate causal effects from the passively observed log data, we control confounds in a matched quasi-experimental design: we identify focal persons who start regularly eating with a fixed partner and match focal persons into pairs such that paired focal persons are nearly identical with respect to covariates measured before acquiring the partner, but the new eating partners diverge with respect to the healthiness of their respective food choice behaviors (before tie formation with the focal person).

We observe that the people who acquire a healthy-eating partner change their habits significantly more toward healthy foods than those acquiring an unhealthy-eating partner. We further identify foods whose purchase frequency is impacted significantly by the partner's healthiness of food choice: coffees and lunch meals are the items that see the largest increase on behalf of those who initiate eating with a healthy-eating partner, whereas the matched counterparts, with the exception of water, increase purchases of items that loosely form a cluster of potentially unhealthy items that should not be consumed in large quantities: soft drinks, drinks from vending machines, condiments, pizza, kebabs, and cr\^epes.

\new{Our findings show that digital traces can be used as a valuable tool for monitoring dietary habits, and can provide valuable insights into the effects of social ties on dietary choices. This work establishes the feasibility of relying on transactional logs in order to monitor food consumption and derive meaningful insights about behavioral patterns taking place at a population scale. Such digital traces can complement small-scale field experiments, making it possible to observe large populations over long time periods.}

\new{By relying on a novel transactional dataset and using carefully designed quasi-experimental methodology, we confirm theories of social influence on food choice postulating that social influence plays a prominent role and has a powerful effect on food intake ~\cite{hetherington2006situational,higgs2016social,shepherd_1999,collins2019two,mollen2013healthy}. Conforming to the behavior of others is adaptive, and individuals find it rewarding~\cite{higgs2016social}. Hence, dietary choices are expected to converge with those of our close social connections.}

\new{Additionally, eating norms are known to reduce the intake of unhealthy foods~\cite{robinson_blissett_higgs_2013} in specific contexts. For example, exposure to social eating norms is known to result in a reduction in the weight of consumed high-energy snack foods~\cite{robinson_blissett_higgs_2013}. The fact that persons exposed to a partner eating healthy foods eat fewer items that should not be eaten in large quantities (soft drinks, drinks from vending machines, condiments, pizza, kebabs, and cr\^epes) compared to the counterparts indicates the presence of such positive influence of others.
}

\new{Finally, we contribute to the rich literature about social influences in eating by demonstrating that, beyond studying social modeling in specific situations~\cite{vartanian2015modeling}, the naturally or experimentally occurring event of tie formation is an important dimension to consider in order to understand the mechanisms of social influence and their potential for influence and interventions.}

\xhdr{Implications} \new{ The most imminent utility of measuring the impact of social tie formation on food choice relying on large-scale passively sensed transaction signals lies in its potential to inform public health interventions on campuses. Designing large-scale nutritional interventions is challenging and logistically complicated. Additionally, it is difficult to predict their impact through experimentation due to ethical concerns stemming from the danger of eliciting undesirable effects. This work shows how observational insights based on passively sensed data can be used to evaluate the impact of potential interventions by estimating the impact of similar interventions that occur naturally, without external experimentation.}

\new{For instance, in order to incentivize healthy eating habits, university or corporate stakeholders might consider launching programs with disclosure and consent that help students or staff connect onsite with ``lunch buddies'' and incentivize consumption of meals in a company. Such programs leveraging peer-led nutritional interventions \cite{yip2016peer} would need to take into consideration self-selective disclosure and consent as confounding factors. 
First, by relying on passively sensed observational data and evaluating the impact of similar interventions, it would be possible to anticipate the impact on the involved individuals ahead of implementing any interventions in the real world. Second, it would be possible to optimize the pairing of people, by estimating on whom the effect of tie formation with others could be strongest. Finally, being informed about what products are most likely to be purchased in modified quantities after tie formation, the stakeholders could optimize the offering of products.}

\new{The question of whether the estimated impact of such naturally occurring interventions is identical to the true causal effect, and whether it is expected to mirror the impact of intentional externally induced interventions, remains. In what follows, we discuss the assumptions that would need to hold, and the limitations that should be considered.
}


\xhdr{Causal assumptions}
We discuss what assumptions are necessary for our study design to let us isolate the causal effect of social-tie formation on food consumption. In particular, we consider the assumptions of the potential\hyp outcomes framework \cite{rubin2005causal} and the extent to which they can be assumed to hold in the present work.

\textit{Stable unit treatment value.}\hspace{1mm} Units are assumed not to interfere with each other. In other words, the treatment assignment of one unit does not affect the outcome of another unit, \ie, there is no ``spillover'' or ``contagion.''  Recall that we require initiation with no more than one peer during the monitored pre- and post-treatment periods. While this restriction ensures that there are no spillovers, our study is unable to capture complex network interactions occurring in on-campus settings. Future work should therefore generalize beyond the studied setup by identifying time-varying treatments and dynamic treatment regimes using g-formula methods \cite{young2011comparative,taubman2009intervening}.

\textit{Consistency.}\hspace{1mm} The potential outcome under a treatment is assumed to be equal to the observed outcome when the actual treatment is received. In other words, the counterfactual outcome for treated units is the observed outcome for controls. While our study design attempts to compare people who are similar up to the moment of receiving the treatment by modeling the propensity to be treated, the assumption that the outcome of the matched person would be exactly the outcome of the treated person is untestable.

\textit{Positivity.}\hspace{1mm} The probability of every treatment for every set of covariates is assumed to be non-zero. In our study, it is reasonable to assume there are no people who could never possibly receive a given treatment.

\textit{Ignorability.}\hspace{1mm} Finally, the ignorability assumption refers to the absence of unmeasured confounding. By performing a sensitivity analysis, we have attempted to assess the possible impact of unobserved biases if the ignorable treatment assignment assumption is violated. This analysis leads us to conclude that our findings are insensitive to small biases.

\xhdr{Limitations} This study is subject to certain limitations, some of which suggest promising directions for future work. Inherent to observational studies, we recognize the inability to infer true causality.
Controlling for all the possible confounds is fundamentally infeasible.
Still, we make a step towards understanding the effects of a certain ``treatment'' on food consumption by
developing a quasi-experimental design based on propensity-score matching and difference-in-differences methods,
whereby we seek to minimize biases due to \textit{observed} confounding variables,
enriched with a quantification of the danger of \textit{unobserved} confounding variables via sensitivity analysis.
Our work, therefore, provides insights based on passively sensed behavioral signals that go far beyond simpler correlational analyses.

We note that the inference of social ties might be imperfect. However, the fact that a large fraction of ties forms precisely at the beginning of the academic year (with the fall semester), when students are exposed to new fellow students (Figure~\ref{fig:2}), and the fact that there is a correlation with ground-truth team membership points towards reliability.
In addition, we note that we might potentially be detecting the onset of co-eating with a delay (\ie, it actually occurs earlier than detected) if peers eat together as part of a larger group and are not directly adjacent in the queue, or if they use cash for the transaction.
That said, we note that any potential delay in estimating the onset would lead to more conservative estimates of the effect of partnering up, as potential changes in the patterns would be counted as purchases before the tie formed.

\textit{Construct validity.}\hspace{1mm} We also consider the issue of the construct validity of our study design, \ie, the degree to which the obtained indirect measurements (transaction logs of food consumption) are reflecting the true phenomenon that is intended to be measured (actual food consumption). It is reasonable to assume that students and staff indeed consume the food that they purchase. However, one cannot eliminate the possibility of persons borrowing the card, or paying for items consumed by other people.
Conversely, people on campus may consume food not recorded in the purchase logs (\eg, food purchased using cash, or food prepared at home or in off-campus restaurants).
Future work should determine the extent to which the assumption that purchasing implies consumption, and vice versa, holds.

\textit{External validity.}\hspace{1mm} Future work should also determine external validity, that is, to what extent behaviors measured on campus reflect other settings off campus, \eg, food consumption---and the complex interplay with food consumption of their social ties at home---in families or in restaurants. People who primarily eat home-cooked food might present a skewed representation of their diet in the studied cafeteria purchase logs, and the fact that individuals might consume food prepared at home introduces unobserved variables into our analyses. For example, people who do not have regular eating partners in extended periods might eat alone because they prefer to eat a peculiar type of food that they cook at home and that is not served on campus, or measured via purchase logs.

\xhdr{Ethical considerations}
Nutrition is a potentially sensitive personal behavior.
To protect user privacy, the log data used here was
accessed exclusively by EPFL personnel involved in this project,
and stored and processed exclusively on EPFL servers.
The data was obtained with approval from EPFL's Data Protection Officer and was anonymized before it was made available to the researchers for analysis.
Finally, we note that our work was conducted retroactively on data that had been collected passively in order to support campus operations. Thus, our analysis did not influence users in any way.

\xhdr{Future work}
This study opens the door for future research directions and potential follow-up studies of the mechanisms of social influence beyond those described, such as the role of the campus-wide sustainability challenge (\cf\ Section~\ref{sec:Transaction log data}), and if eating with a healthy-eating partner is correlated with other behavioral changes (\eg, switching cafeterias).
 Future work should further understand the measured behavioral changes by understanding the underlying mechanisms and the social contexts in which the foods are consumed in modified quantities (\eg, socializing by meeting for a cup of coffee versus consuming a meal together). Future work should demonstrate whether modified habits are retained in those individuals who return later to eating alone for long stretches, or, if habits are not retained, how the individual regresses to the state they were in before the tie formation, and if so, what is the rate of the decay back.

\xhdr{Beyond food}
To the best of our knowledge, this is the first study to leverage large-scale transactional data to retrospectively evaluate the impact of implicit dietary behavioral interventions. We show that purchase logs are useful as an effective sensor to detect behavioral changes in individuals who interact with others. Beyond food purchase logs on campus, this work highlights how careful quasi-experimental comparisons leveraging passively sensed data can be used to measure complex interactions between individuals. The methodology of isolating influence through matched pairwise comparisons has the potential to allow for detecting behavioral changes also in areas relevant to other social computing scenarios. Ambitious questions of how individuals impact behaviors or skills of one another in the context of purchasing decisions, exercise habits, coaching or counselling, travelling, sleep patterns, online interactions, collaborative writing, or coding practice could all be tackled by identifying the onset of such interactions, and comparing outcomes in similar matched persons while controlling for confounds.

\xhdr{Code and data}
Although the transaction logs cannot be shared publicly,
for transparency we make all analysis code, as well as the food-item and healthiness-score annotations, publicly available:
\url{https://github.com/epfl-dlab/tie-formation-food}

\begin{acks}
We would like to thank
Nils Rinaldi, Aurore Nembrini, and Philippe Vollichard for their help in obtaining and anonymizing the data.
We are also grateful to Jonas Racine and Kiran Garimella for early help with data engineering,
and to our reviewers for their helpful suggestions.
This project was funded by the Microsoft Swiss Joint Research Center.
\end{acks}

\bibliographystyle{ACM-Reference-Format}
\bibliography{bibliography}

\end{document}
\endinput